\documentclass[11pt]{article}
\pdfoutput=1
\usepackage{jheppub}
\usepackage[T1]{fontenc}
\usepackage[utf8]{inputenc}
\usepackage{epsfig}
\usepackage{graphicx}
\usepackage{latexsym}
\usepackage{textcomp}
\usepackage{amssymb}
\usepackage{amsfonts,amsthm,amstext,amscd}
\usepackage{amsmath}
\newcommand{\be}{\begin{equation}}
\newcommand{\ee}{\end{equation}}
\newcommand{\beq}{\begin{equation}}
\newcommand{\eeq}{\end{equation}}
\newcommand{\bea}{\begin{eqnarray}}
\newcommand{\eea}{\end{eqnarray}}

\newcommand{\bear}{\begin{eqnarray}}
\newcommand{\eear}{\end{eqnarray}}

\newcommand{\morder}[1]{\mathcal{O}\left(#1\right)}
\newbox\pippobox

\def\II{\relax{\rm I\kern-.18em I}}

\def\e{\epsilon}
\def\m{\mu}
\def\n{\nu}

\def\pa{\partial}

\def\tr{\ensuremath{\mathrm{Tr}}}

\def\l{\lambda}

\def\gf{w}
\def\h{\kappa}

\title{Cool baryon and quark matter in holographic QCD}
\author[\dag]{Takaaki Ishii,}
\author[\ddag]{Matti J\"arvinen}
\author[\ddag]{and Govert Nijs}
\affiliation[\dag]{Department of Physics, Kyoto University, Kyoto 606-8502, Japan}
\affiliation[\ddag]{Institute for Theoretical Physics and Center for Extreme Matter and Emergent Phenomena, Utrecht University, 3584 CE Utrecht, The Netherlands}
\emailAdd{ishiitk@gauge.scphys.kyoto-u.ac.jp}
\emailAdd{m.o.jarvinen@uu.nl}
\emailAdd{g.h.nijs@uu.nl}
\abstract{We establish a holographic bottom-up model which covers both the baryonic and quark matter phases in cold and dense QCD. This is obtained by including the baryons using simple approximation schemes in the V-QCD model, which also includes the backreaction of the quark matter to the dynamics of pure Yang-Mills. 
We examine two approaches for homogeneous baryon matter: baryons as a thin layer of noninteracting matter in the holographic bulk, and baryons with a homogeneous bulk gauge field. We find that  
the second approach exhibits phenomenologically reasonable features.  
At zero temperature, the vacuum, baryon, and quark matter phases are separated by strongly first order transitions as the chemical potential varies. 
The equation of state in the baryonic phase is found to be stiff, i.e., the speed of sound clearly exceeds the value $c_s^2=1/3$ of conformal plasmas  at high baryon densities. }
\preprint{KUNS-2752}

\begin{document}
\maketitle
\flushbottom

\section{Introduction and summary}

Studying dense matter in QCD has turned out to be a hard problem with 
many unresolved questions remaining~\cite{Brambilla:2014jmp}. 
The main theoretical tool, perturbation theory, applies only to asymptotically high densities where QCD becomes a free theory~\cite{Kurkela:2009gj}. Effective methods such as chiral perturbation theory are useful to describe nuclear matter at low densities~\cite{Machleidt:2011zz,Tews:2012fj}. But the ranges where  
these methods can be trusted leave a wide gap at intermediate densities where reliable and accurate approaches are not available. Moreover, first principles lattice field theory methods only work at small values of the baryon chemical potential due to the famous sign problem, and cannot be used to study properties of cold QCD matter. Consequently, even basic observables such as the equation of state of cold matter have significant uncertainties after applying known theoretical and experimental constraints (see, e.g.,~\cite{Annala:2017llu}). The intermediate densities, where uncertainties are at their largest, are physically relevant: they contain the phase transition (or possibly a crossover, or several transitions) between the nuclear matter and quark matter phases, and the densities of neutron star cores are known to lie within this region.

In the absence of applicable first-principles methods, model computations can give useful information about the properties of strongly interacting QCD matter in the regime of the transition between the baryon and quark matter phases. In this article, we will study this regime by using gauge/gravity duality. One of the weaknesses in this approach  
is that no exact gravity dual for QCD is known, and typically the models available in the literature have similar features as QCD but fail to reproduce in detail  the thermodynamics of QCD, for example. 
However, recently progress towards more realistic and reliable modeling of QCD has been made, which motivates us to apply these models to cold and dense QCD matter.  We will use one of the most realistic holographic models available (V-QCD). 

V-QCD is a class of holographic models for QCD, obtained through a fusion~\cite{Jarvinen:2011qe} of two frameworks: improved holographic QCD (IHQCD)~\cite{Gursoy:2007cb,Gursoy:2007er,Gursoy:2008bu,Gursoy:2008za,Gursoy:2009jd} for the gluon sector, and a setup based on Sen-like tachyonic Dirac-Born-Infeld (DBI) actions for the quark sector~\cite{Bigazzi:2005md,Casero:2007ae}. The former framework is inspired by five dimensional noncritical string theory, and the latter is obtained by introducing a pair of space filling branes in this background. In the Veneziano limit where one takes both the number of colors and number of flavors to infinity keeping their ratio $x_f \equiv N_f/N_c$ fixed, the two sectors are fully backreacted as one expects for ordinary QCD (with three colors and 2-3 light flavors). The model is not derived from string theory strictly: in the end one switches to bottom-up approach because on the one hand the results do not match precisely with known QCD phenomenology, and on the other hand the stringy derivation cannot be made exact (in particular due to working in the Veneziano limit). Therefore one generalizes the action to contain certain potential functions, which are then chosen to agree with qualitative QCD features and/or fitted to lattice and experimental data, in a rough analogy to effective field theory.

The thermodynamics of V-QCD has been studied in earlier work~\cite{Alho:2012mh,Alho:2013hsa,Alho:2015zua} and shown to agree qualitatively with several known properties of QCD, such as the main features of the phase diagram as a function of temperature and chemical potential. After comparison with lattice data for the equation of state at small chemical potential, the model was shown to produce an equation of state for cold and dense quark matter which 
agrees with known experimental and theoretical constraints for QCD~\cite{Jokela:2018ers}. A remaining major task in order to establish a model including the basic features of cold QCD is the inclusion of baryon physics in V-QCD. In this article, we will take the first steps in this direction. 

Baryons are introduced as solitonic configurations in holographic models for large-$N_c$ QCD. In top-down construction, a D-brane joining $N_c$ open strings gives a baryon vertex \cite{Witten:1998xy,Gross:1998gk} and provides a baryon number through Chern-Simons (CS) terms. Following this notion, baryons have been considered in effective holographic theories, especially in the Witten-Sakai-Sugimoto (WSS) model \cite{Witten:1998zw,Sakai:2004cn,Sakai:2005yt}. First, an approximation by a small size instanton was used to introduce a baryon \cite{Hong:2007kx,Hata:2007mb,Hong:2007dq,Hashimoto:2008zw,Kim:2008pw}, and then this has been generalized to include contributions beyond the instanton approximation \cite{Cherman:2009gb,Cherman:2011ve,Bolognesi:2013nja,Rozali:2013fna}. A baryon solution has been constructed also as in bottom-up AdS/QCD \cite{Pomarol:2008aa,Panico:2008it}. Multi-baryon solutions are also studied in the WSS model \cite{Kaplunovsky:2012gb,deBoer:2012ij,Kaplunovsky:2015zsa,Preis:2016fsp,BitaghsirFadafan:2018uzs}. 
For dense baryonic matter in the QCD phase diagram, homogeneous approximations have been utilized in the WSS model \cite{Bergman:2007wp,Rozali:2007rx,Ghoroku:2012am,Li:2015uea,Elliot-Ripley:2016uwb} as well as another approach based on probe branes \cite{Gwak:2012ht,Evans:2012cx}. With these approximations, baryonic matter phases can be realized in the low temperature high-density region in the phase diagram.

In this article, we carry out the first study of baryons in V-QCD. We restrict to approximations where the baryon configurations are homogeneous in spatial directions and which hence simplifies the analysis considerably. We work in an isospin symmetric setup and also neglect the effects due to light quark masses.
We adopt two approaches:
\begin{itemize}
 \item The first, given in Sec.~\ref{sec:pointlike}, is to introduce a nondynamical thin layer of baryons localized in the holographic coordinate. This approach is essentially equivalent to treating the baryons as point-like sources, which is the picture arising in the WSS model at large coupling \cite{Bergman:2007wp,Rozali:2007rx}\@.  
\item The second, slightly more advanced method given in Sec.~\ref{sec:homogeneous}, employs a homogeneous ansatz with $SU(2)$ flavor symmetry for the spatial components of the non-Abelian flavor gauge field, sourcing baryon density through the CS coupling.  
The region where the solution is highly inhomogeneous in the bulk is modeled as a discontinuity of the homogeneous baryon field.  
\end{itemize} 
We summarize the main results from both these approaches in the following. The main message is that the first method works 
unsatisfactorily, but the second one 
provides reasonable results.

First, we consider the approach  with a thin layer of baryons of Sec.~\ref{sec:pointlike}. We also include the full backreaction of this layer to the five dimensional gravitational background. We make the following observations:
\begin{itemize}
 \item Stabilizing the layer of baryon matter turns out to be hard as it has the tendency of decaying by falling in the IR (deep in the bulk) where the approximations made in this approach also break down. We can choose the potentials such that the layer stays near the boundary of the five dimensional space, but as it turns out, this leads to another problem: the obtained phase diagram (even in the absence of baryons) is at odds with QCD phenomenology. In particular, confinement can be obtained only at very small chemical potentials.
 \item If we choose the action so that the baryons are present, baryonic phases appear in the expected region of the phase diagram: at low temperatures and intermediate chemical potentials. Consequently, we obtain phases where the charge is sourced in part by the baryons and in part by quark matter. We, however, also obtain a chirally symmetric baryonic phase which looks exotic from a QCD intuition. The phase structure obtained in this approach is summarized in Fig.~\ref{fig:phases}\@.
\end{itemize}
It is apparent that the problems observed within this approach are weaknesses of the approach rather than the V-QCD model. This motivates us to consider an improved method.

We discuss the setup with a homogeneous non-Abelian bulk field in Sec.~\ref{sec:homogeneous}, and demonstrate that this indeed improves the results of Sec.~\ref{sec:pointlike} in several ways. We  consider this baryon ansatz in the probe limit, i.e., on top of a fixed gravitational background. The main results from this approach, and therefore the main results of this article, are the following:
\begin{itemize}
 \item The approach is seen to capture the coupling of the baryons to another bulk field (the tachyon) which is dual to the $\bar qq$ operator and therefore controls chiral symmetry and its breaking. It is the coupling to the tachyon, which was missing in the simpler approach of Sec.~\ref{sec:pointlike}, that prevents the baryons from falling in the IR.
 \item The phase diagram has the expected structure: baryons dominate at low temperatures and intermediate chemical potentials, between the confined vacuum phase (dominant at low chemical potentials) and deconfined quark gluon plasma (QGP) phase (dominant at high chemical potentials). All phase transitions between these phases are of first order. The phase diagram for this approach is shown in Fig.~\ref{fig:smearedphases}\@. Notice that the thermodynamics in the vacuum and in the baryon phase is independent of the temperature, but nontrivial temperature dependence is included in the QGP phase.
 \item As the density is increased, the equation of state in the baryonic phase becomes stiff, and the speed of sound rises well above the conformal value $c_s=1/\sqrt{3}$. This is interesting because with stiff equations of state it is easier to pass the constraints set by observations of masses and deformability of neutron stars. The basic picture is therefore the following: the nuclear (quark) matter has a stiff (soft) equation of state, and the latent heat at the baryon to QGP transition is sizable. This agrees with the earlier analysis in V-QCD~\cite{Jokela:2018ers} where polytropic interpolations were used to model the baryonic phase.
\end{itemize}
Readers interested in these main results can safely skip Sec.~\ref{sec:pointlike}, as the discussion of Sec.~\ref{sec:homogeneous} can be followed independently.

Another important result arises as a by-product of the problem found in Sec.~\ref{sec:pointlike}. Namely, having the correct confinement properties of the phase diagram sets a previously unknown constraint to the V-QCD models. This constraint is actually completely independent of baryon physics. That is, the coupling of the gauge fields in the DBI action of the flavor sector (function $w(\l)$ defined below) is constrained to agree (up to small corrections) the IR behavior predicted by string theory, which complements similar results for the other coupling functions of the model found in the literature~\cite{Gursoy:2007cb,Gursoy:2007er,Jarvinen:2011qe,Arean:2012mq,Arean:2013tja,Arean:2016hcs}. This is discussed in detail in Appendix~\ref{app:wasympt}.

The paper is organized as follows.
First, in Sec.~\ref{sec:VQCD}, we introduce the V-QCD model.
In particular, we work out the CS terms, which are essential for computing the baryon physics.
In sections \ref{sec:pointlike} and \ref{sec:homogeneous}, we consider the thin layer and homogeneous non-Abelian approaches for the baryons, respectively. Discussion and outlook are given in Sec.~\ref{sec:conclusions}. Additional details on the computations are given in the appendices.

\section{V-QCD and setup for baryon physics}\label{sec:VQCD}

\subsection{The holographic action}

We start by reviewing the two basic building blocks of V-QCD. 
First, improved holographic QCD~\cite{Gursoy:2007cb,Gursoy:2007er,Gursoy:2008bu,Gursoy:2008za,Gursoy:2009jd} gives the description of the dynamics of gluons. It is a bottom-up model for pure Yang-Mills motivated by noncritical string theory. Second, the flavor sector is introduced through a tachyonic DBI action, inspired by a space filling $D4 - \overline{D4}$ configuration~\cite{Bigazzi:2005md,Casero:2007ae}.
The two sectors are fully backreacted in the Veneziano limit: 
\be
 N_c \to \infty\,,\qquad N_f \to \infty \,,\qquad N_f/N_c \equiv x_f \quad \mathrm{fixed}\,,\qquad g^2N_c \quad \mathrm{fixed}\,.
\ee
Such backreacted models (V-QCD) were constructed in~\cite{Jarvinen:2011qe}, and these are the models we discuss in this article. A similar setup was considered in the probe limit in~\cite{Iatrakis:2010zf,Iatrakis:2010jb}. 

The relevant part of the dictionary is the following:
\begin{itemize}
 \item The dilaton field $\lambda=e^\phi$ is dual to the $\mathrm{Tr}\ F^2$ operator and therefore sources the 't Hooft coupling in Yang-Mills theory. This is the only field from the IHQCD sector which we will consider in this article.
 \item The tachyon field $T^{ij}$ is dual\footnote{More precisely, the duality is defined through the boundary Lagrangian  $ \propto \bar q  T (1+\gamma_5) q/2 + \bar q T^\dagger (1-\gamma_5) q/2$~\cite{Arean:2016hcs}.} to the quark bilinear $\bar q^i q^j$ and sources the quark mass matrix. It arises from the strings stretching between the $D4$ and $\overline{D4}$ branes.
 \item The left and right handed gauge fields $\left(A_{L/R}^\mu\right)^{ij}$ living on the branes are dual to the left and right handed currents $\bar q^i \gamma^\mu(1 \pm \gamma_5) q^j/2$.
\end{itemize}

The action of the full model consist of several terms:\footnote{For finite temperature studies one must also include the appropriate Gibbons-Hawking term which will be specified below.}
\be
 S_\textrm{V-QCD} = S_\mathrm{glue} + S_\mathrm{DBI} + S_\mathrm{CS} \ ,
\ee
where the first term is the action of IHQCD and the other two terms describe the dynamics of the flavor branes. We will first discuss the first two terms, and the relevant pieces of the CS action $S_\mathrm{CS}$ will be given in Sec.~\ref{sec:CS}.

The action for the gluon dynamics is given by 
\be \label{glueaction}
 S_\mathrm{glue} = M^3N_c^2 \int d^5x\, \sqrt{-\mathrm{det}\, g} \left[R-\frac{4}{3}\frac{(\partial \l)^2}{\l^2}+V_g(\l)\right] \,,
\ee
where $M$ is the five dimensional Planck mass. The dilaton potential will be chosen appropriately to mimic the physics of QCD.

The full flavored DBI action of the model reads
\be
 S_\mathrm{DBI} =  - \frac{1}{2} M^3 N_c\,  {\mathbb Tr} \int d^5x\,
\left(V_f(\l,T^\dagger T)\sqrt{-\det {\bf A}^{(L)}}+V_f(\l, TT^\dagger)\sqrt{-\det {\bf A}^{(R)}}\right)\,,
\label{generalact}
\ee
with the radicands defined through
\begin{align}
{\bf A}_{MN}^{(L)} &=g_{MN} + \gf(\l,T) F^{(L)}_{MN}
+ {\h(\l,T) \over 2 } \left[(D_M T)^\dagger (D_N T)+
(D_N T)^\dagger (D_M T)\right] \,,\nonumber\\
{\bf A}_{MN}^{(R)} &=g_{MN} + \gf(\l,T) F^{(R)}_{MN}
+ {\h(\l,T) \over 2 } \left[(D_M T) (D_N T)^\dagger+
(D_N T) (D_M T)^\dagger\right] \,,
\label{Senaction}
\end{align}
and the covariant derivative given by
\be
D_M T = \partial_M T + i  T A_M^L- i A_M^R T\,.
\ee
Our convention for the field strengths is such that
\be
 F^{(L/R)} = dA_{L/R} - i A_{L/R} \wedge A_{L/R}\,.
\ee
The fields  $A_{L}$, $A_{R}$ and $T$ are $N_f \times N_f$ matrices in the flavor space, and $\mathbb{T}r$ denotes the trace over flavor indices. Notice that the full non-Abelian DBI action is not known, and typically a symmetrized trace prescription~\cite{Tseytlin:1997csa}  is assumed. The first few corrections as a series in $F$ are know precisely~\cite{Refolli:2001df,Koerber:2001uu,Grasso:2002wb,Keurentjes:2004tu}. In this article we will only consider non-Abelian configurations using the first nontrivial term in the expansion on top of an Abelian background, in which case ambiguities in the prescription are absent, and it is enough to use a standard trace in~\eqref{generalact}.
Under the left and right
$U(N_f)$ gauge transformations the fields transform as
\begin{align}
&A_L\to V_L\,A_L\,V_L^\dagger-idV_L\,V_L^\dagger\,,\qquad
A_R\to V_R\,A_R\,V_R^\dagger-idV_R\,V_R^\dagger\,,\nonumber\\
&T\to V_R\,T\,V_L^\dagger\,,\hspace{3.1cm}
T^\dagger\to V_L\,T^\dagger\,V_R^\dagger\,,
\label{gaugetransf}
\end{align}
with $V_L\,V_L^\dagger=\mathbb{I}_{N_f} =V_R\,V_R^\dagger$.

We will make the simplifying assumption that the couplings $\gf$ and $\h$ depend on $\l$ only. Moreover we consider backgrounds where the tachyon is flavor independent:
\be
T=\tau(r)\,\mathbb{I}_{N_f}\,,
\label{comptach}
\ee
and use a Sen-like tachyon potential
\be
V_f(\l,TT^\dagger)=V_{f0}(\l) e^{- a \tau^2} \,.
\label{tachpot}
\ee
where we take $a$ to be a constant. Its value can be absorbed into the normalization of the tachyon, so we will set $a=1$ from now on.
As for the metric, 
our ansatz reads
\be
ds^2=e^{2 A(r)} (-f(r)dt^2+d\mathbf{x}^2+f(r)^{-1}dr^2)\,.
\label{bame}
\ee

In order to determine the model completely, one needs to also specify the potential functions $V_g(\l)$, $V_{f0}(\l)$, $\kappa(\l)$, and $w(\l)$. The general idea is the following: the asymptotics of most of these functions both in the UV ($\l \to 0$) and in the IR ($\l \to \infty$) are tightly constrained by agreement with QCD. In more detail, constraints, e.g., from confinement~\cite{Gursoy:2007cb,Gursoy:2007er}, consistency of the backgrounds, linear glueball/meson trajectories~\cite{Gursoy:2007er,Jarvinen:2011qe,Arean:2013tja}, and regularity of the model at finite $\theta$-angle~\cite{Arean:2016hcs} set the power and in some cases the subleading logarithmic term of the IR behavior for the functions (for example $V_g \sim \l^{4/3}\sqrt{\log \l}$ as $\l \to \infty$). 

In the UV, i.e., in the weak coupling regime, holographic models are generally not reliable. However, to set the best boundary conditions for the IR physics, we choose the UV behavior of the functions to agree with QCD perturbation theory: as usual we require that the correct UV dimensions of the various operators are reproduced, but in addition we require agreement with asymptotic freedom~\cite{Gursoy:2007cb,Gursoy:2007er}, with RG flow of the quark mass~\cite{Jarvinen:2011qe}, and behavior at large quark mass~\cite{Jarvinen:2015ofa}. Interestingly, this is obtained if all the functions go to constants in the UV with perturbative corrections in $\l$\@.

In the intermediate region, $\l = \mathcal{O}(1)$, the remaining degrees of freedom in the potentials need to be fitted to QCD data from experiments and from lattice computations. This has been considered for IHQCD in~\cite{Gursoy:2009jd} and started for full V-QCD in~\cite{Jokela:2018ers}.

For the baryon physics  a particularly important choice is that of the function $w(\l)$. We will discuss the choice in more detail below. The explicit choices of potentials which we use in this article are given in Appendix~\ref{app:potentials}.

\subsection{Thermodynamics in the absence of baryons}\label{sec:backgroundthermo}

We first discuss the physics and the phase diagram in the absence of baryons. Then we only have a vectorial flavor singlet gauge field $A_L=A_R = \mathbb{I}_{N_f} \Phi(r) dt$ giving nonzero charge density and chemical potential.
Inserting the expressions for the fields and the potentials the DBI action evaluates to 
\be
 S_\mathrm{DBI}^{(0)} =  -  M^3 N_c N_f  \int d^5x\, V_{f0}(\l)e^{-\tau^2}
\sqrt{-\det g}\sqrt{1+ e^{-2A}f \kappa(\l)(\tau')^2-e^{-4A}w(\l)^2(\Phi')^2}\,.
\label{SDBI0}
\ee
The thermodynamics of IHQCD has been studied in~\cite{Gursoy:2008bu,Gursoy:2008za}, and thermodynamics in the V-QCD setup has been discussed at zero $\mu$ in~\cite{Alho:2012mh,Alho:2015zua} and at nonzero $\mu$ in~\cite{Alho:2013hsa,Jokela:2018ers} at zero quark mass. Quark mass effects have been considered in~\cite{Jarvinen:2015ofa}. The V-QCD action has two classes of solutions which either have a horizon or not. The ``thermal gas'' solutions without a horizon extend from the UV boundary to the IR singularity which is of the ``good'' kind~\cite{Gubser:2000nd}. They have trivial thermodynamics: the pressure is zero, whereas the pressure is nontrivial and $\mathcal{O}(N_c^2)$ for the black hole solutions with a horizon. The different scalings of the pressure with $N_c^2$ can be interpreted as an order parameter for confinement~\cite{Alho:2012mh,Alho:2015zua}. 

Both thermal gas (TG) and black hole (BH) solutions have two further variants which either have or do not have  a scalar tachyon hair, i.e., nonzero bulk condensate of the field $\tau$. The (non)existence of tachyon hair determines through the dictionary whether chiral symmetry is broken or not. Therefore there are in total four phases. Studies with choices of the potentials $V_g$, $V_f$, $\kappa$, and $w$ that reproduce various features of QCD~\cite{Alho:2012mh,Alho:2013hsa,Jokela:2018ers} have shown that three of these solutions may be dominant for values of $x_f = \morder{1}$ relevant for ordinary QCD:
\begin{itemize}
 \item The tachyonic thermal gas solution, which is identified as the description of the confined and chirally broken phase in QCD. This phase appears at low values of the temperature and chemical potential, i.e., for $T \lesssim \Lambda$ and  $\mu \lesssim \Lambda$ where $\Lambda$ is the characteristic energy scale of the model which we will define precisely below in Sec.~\ref{sec:pointlikeeom}. 
 \item The tachyonless black hole solutions, which are identified as the chirally symmetric deconfined quark matter phase in QCD. This phase dominates for large values of temperature or the chemical potential.
 \item Tachyonic black hole solutions, which describe an intermediate deconfined but chirally broken phase. Depending on the precise choice of the potentials this phase may or may not be present in the phase diagram. As it turns out, fits to lattice data disfavor its existence~\cite{Jokela:2018ers}. For the potentials which will be used in this article, it is subdominant for all values of $T$ and $\mu$ and therefore does not appear in the phase diagram.
\end{itemize}

The confinement/deconfinement phase transition, which is realized as a Hawking-Page phase transition is always first order. It is possible that stringy loop corrections, which we shall not consider in this article, turn the first order transition into a higher order transition or a crossover at low values of $\mu$~\cite{Alho:2015zua}. The chiral transition may also be of second order, if the intermediate phase exists so that it is separated from the confinement/deconfinement transition, but otherwise it is of first order. A rather similar phase structure has been found also in models based on a D3/D7 brane system, see~\cite{Evans:2010iy,Evans:2010hi}.

As it turns out, requiring the phase diagram to have the desired phase structure leads to a new nontrivial constraint for the potentials. More precisely, requiring that the TG phase extends to nonzero $\mu$ constrains the IR asymptotics of the function $w(\l)$. We discuss this in detail in Appendix~\ref{app:wasympt}. The result is that in the IR asymptotics, $w(\l) \sim \l^{-w_p}$ as $\l \to \infty$, we must have $w_p \ge 4/3$. Consideration of the meson spectra (in particular the splitting between vector and axial vector mesons) sets $w_p \le 4/3$~\cite{Arean:2013tja}, which pins down $w_p=4/3$ as the only remaining possibility. This choice was indeed used in~\cite{Alho:2013hsa,Jokela:2018ers}. The result complements earlier findings for the leading IR asymptotics as follows: Results on confinement and on glueball trajectories fix $V_g \sim \l^{4/3}$~\cite{Gursoy:2007cb,Gursoy:2007er}. Meson trajectories and regularity of chirally broken solutions set $\kappa \sim  \l^{-4/3}$~\cite{Arean:2013tja}. Regularity of solutions with chiral symmetry and at finite $\theta$-angle, and agreement with QCD phase diagram as a function of $x_f = N_f/N_c$ require $4/3 \le v_p \le 10/3$ in $V_{f0} \sim \l^{v_p}$. Moreover, complete regular solutions could be found numerically only for $v_p \lesssim 3$~\cite{Jarvinen:2011qe,Arean:2016hcs}. These results hold up to logarithmic terms in $\l$ (which are also fixed for $V_g$ and $\kappa$~\cite{Gursoy:2007cb,Gursoy:2007er,Arean:2013tja}). Interestingly, after including the result of Appendix~\ref{app:wasympt}, the power laws for $V_g$, $\kappa$, and $w$, determined by matching with QCD agree with expectation from noncritical string theory in the Einstein frame (even though this was not required in the fit), see~\cite{Arean:2013tja}. In addition, the string theory result $v_p=7/3$ also lies in the acceptable range.

The detailed comparison of the full V-QCD model with lattice data was initiated in~\cite{Jokela:2018ers}. This was done by fitting the thermodynamics of the model in the deconfined, chirally symmetric phase to lattice data near $\mu=0$ in full QCD with 2+1 flavors. One of the main results after the fit was, that the intermediate phase with deconfinement and broken chiral symmetry, was absent (as we already mentioned above). Another important result was that the extrapolated equation of state (EoS) in the deconfined QGP phase from $\mu=0$ and finite $T$ to $T=0$ and finite $\mu$ was typically in agreement with theoretical and observational bounds. The backgrounds resulting from this fit will be the starting point for this article, on top of which the baryon dynamics will be added. 
Extrapolated EoSs for QCD on the $(\mu,T)$-plane have also been considered earlier using holography~\cite{DeWolfe:2010he,DeWolfe:2011ts,Knaute:2017opk,Critelli:2017oub} and field theory~\cite{Rebhan:2003wn,Peshier:2005pp,Ratti:2005jh}.

\subsection{DBI action for small non-Abelian gauge fields}

Then we include the baryonic terms assuming that the amplitude of the soliton is so small that it can be treated as a small perturbation on top of the background. That is, with slight abuse of notation, we replace $A_{L/R} \to \mathbb{I}_{N_f}\Phi dt + A_{L/R}$ and treat $A_L$ and $A_R$ (but not $\Phi$) as small perturbations. We will specify below what exactly are the leading nontrivial terms in the expansion that we will consider.

This division of the gauge field into $\Phi$ and the flavor singlet part of $A_L+A_R$ 
is however not well-defined for generic baryon fields $A_{L/R}$. For our purposes it is enough to fix this by requiring that the soliton part satisfies
\be \label{conscond}
 \int d^4x\, \mathbb{T}r\left(F^{(L)}_{rt}+F^{(R)}_{rt}\right) = 0 \ .
\ee

We go on developing~\eqref{generalact} as a series at small gauge fields. We note that
\begin{align} \label{ALseries}
 {\bf A}_{MN}^{(L)}  = g_{MN} &+ \kappa(\l)\delta_M^r\delta_N^r(\tau')^2+ w(\l)(\delta_M^r\delta_N^t-\delta_M^t\delta_N^r)\Phi'+   w(\l) F^{(L)}_{MN}&\nonumber\\ &+\frac{\kappa(\l)\tau^2}{2}\left(A_M A_N+A_NA_M\right) \,,&
\end{align}
where $A=A_L-A_R$. A similar identity holds for ${\bf A}^{(R)}$. The last two terms in~\eqref{ALseries} capture the contribution from the soliton and are treated as small perturbations. We define the effective metric as
\be
 \tilde g_{MN} \equiv  g_{MN} + \kappa(\l)\delta_M^r\delta_N^r(\tau')^2+ w(\l)(\delta_M^r\delta_N^t-\delta_M^t\delta_N^r)\Phi' 
\ee
so that
\be
\left( \tilde g^{-1}\right)^{MP} {\bf A}_{PN}^{(L)} = \delta^M_N + w(\l) \left( \tilde g^{-1}\right)^{MP}F^{(L)}_{PN} +\frac{\kappa(\l)\tau^2}{2}\left( \tilde g^{-1}\right)^{MP}\left(A_P A_N+A_NA_P\right) \ .
\ee
Taking the determinant and rearranging the terms, we find
\begin{align} \label{ALexp}
 \sqrt{-\det {\bf A}^{(L)}} \simeq \sqrt{-\det\tilde g}&\Bigg[1 + \frac{w(\l)}{2} \left( \tilde g^{-1}\right)^{MN}F^{(L)}_{NM}+\frac{\kappa(\l)\tau^2}{4}\left( \tilde g^{-1}\right)^{MN}\left(A_N A_M+A_MA_N\right)\nonumber\\
 &+ \frac{w(\l)^2}{8}\left( \left( \tilde g^{-1}\right)^{MN}F^{(L)}_{NM}\right)^2-\frac{w(\l)^2}{4}\left( \tilde g^{-1}\right)^{MN}F^{(L)}_{NP}\left( \tilde g^{-1}\right)^{PQ}F^{(L)}_{QM}\Bigg] 
\end{align}
where we only included the terms up to quadratic order in $F^{(L/R)}$, corresponding to the expansion in $\alpha'$, and the leading nontrivial additional term appearing due to the presence of the background tachyon field. We further notice that 
\be 
\left( \tilde g^{-1}\right)^{MN}F^{(L)}_{NM} = 2 \Xi^{-1} e^{-4A} w(\l)\Phi'F^{(L)}_{rt} \,, 
\ee 
where
\be \label{Rdef} \Xi =\frac{\det \tilde g}{\det g} =1+ e^{-2A}f\kappa(\l)(\tau')^2-e^{-4A}w(\l)^2(\Phi')^2 \ee since only the antisymmetric terms in $\tilde g^{-1}$ contribute. Let us denote by $(\tilde g^{-1})_s$ the remaining diagonal and symmetric terms:
\be
 (\tilde g^{-1})_s = e^{-2A}\,\mathrm{diag}\left(-f^{-1}\Xi^{-1}(1+e^{-2A}f\kappa(\l)(\tau')^2),1,1,1,f \Xi^{-1}\right) \,,
\ee
where the indexes are ordered as in the expressions for the metric above: $(t,x_1,x_2,x_3,r)$.
In the last two terms of~\eqref{ALexp} the contributions from the antisymmetric terms exactly cancel. Putting these observations together,
\begin{align}
 \sqrt{-\det {\bf A}^{(L)}} \simeq \sqrt{-\det\tilde g}\Bigg[1 &+ \Xi^{-1} e^{-4A} w(\l)^2\Phi'F^{(L)}_{rt}+\frac{\kappa(\l)\tau^2}{2}\left( \tilde g^{-1}\right)_s^{MN}A_MA_N &\nonumber\\
 &-\frac{w(\l)^2}{4}\left( \tilde g^{-1}\right)_s^{MN}F^{(L)}_{NP}\left( \tilde g^{-1}\right)_s^{PQ}F^{(L)}_{QM}\Bigg] \,.
\end{align}
We are now ready to write down the leading term of the DBI action in the flavored gauge fields:
\begin{align} \label{SDBI1final}
  S_\mathrm{DBI}^{(1)} &= - M^3 N_c  \int d^5x\, V_{f0}(\l) e^{-\tau^2}
\sqrt{-\det g}\sqrt{\Xi}\Bigg[
\frac{\kappa(\l)\tau^2}{2}\left( \tilde g^{-1}\right)_s^{MN}\,\mathbb{T}r A_MA_N&\nonumber\\
&\phantom{=} -\frac{w(\l)^2}{8}\left( \tilde g^{-1}\right)_s^{MN}\left( \tilde g^{-1}\right)_s^{PQ}\,\mathbb{T}r\left(F^{(L)}_{NP}F^{(L)}_{QM}+F^{(R)}_{NP}F^{(R)}_{QM}\right) \Bigg] \,,
\end{align}
where we also included the terms arising from ${\bf A}^{(R)}$ and used~\eqref{conscond}. Notice that up to quadratic order in the gauge fields the DBI action is unambiguous: the result is independent of the order of the (non-Abelian) fields. For higher order terms a specific prescription (e.g., the symmetrized trace) would need to be chosen.

\subsection{Chern-Simons terms}\label{sec:CS}

The CS terms determine how the solitons source baryonic charge. These terms depend on a CP-odd potential $V_a(\l,\tau)$~\cite{Arean:2016hcs} which must satisfy certain requirements: the normalization in the UV ($\l=0=\tau$) must reproduce the correct axial anomaly and perturbative corrections in $\l$ must vanish due to the perturbative nonrenormalization of the anomaly. 
In principle we could work with a generic CP-odd potential $V_a(\l,\tau)$ but we choose the string motivated ansatz $V_a(\l,\tau)=e^{-b\tau^2}$.   
The inclusion of the constant $b$ reflects the findings of~\cite{Arean:2016hcs}: in order for the model to have regular IR solutions in the presence of a finite $\theta$-angle, the contributions from the CS terms had to vanish faster than those coming from the DBI. The easiest way to arrange this is to take $b>1$ in the CS action. In this section we will set $b=1$ for notational simplicity. It will be reintroduced later by rescaling $\tau$.

We compute here explicitly the coupling of  $\Phi$ to the instanton density arising from these terms. The relevant CS term is given by~\cite{Casero:2007ae}
\be
 S_\mathrm{CS} = \frac{iN_c}{4\pi^2}\int  \Omega_5 \,,
\ee
where
\be \label{Omega5}
\begin{split}
\Omega_5&=\frac{1}{6}\tr \,e^{-\tau^2}\!\left\{ -iA_L
\wedge F^{(L)}\wedge F^{(L)} +\frac{1}{2}A_L\wedge A_L \wedge
A_L \wedge F^{(L)} +\frac{i}{10} A_L\wedge A_L \wedge A_L
\wedge A_L\wedge A_L\right.\\
&+iA_R\wedge F^{(R)}\wedge F^{(R)} -\frac{1}{2}A_R\wedge A_R
\wedge A_R \wedge F^{(R)} -\frac{i}{10} A_R\wedge A_R
\wedge A_R\wedge A_R\wedge A_R\\
&+\tau^2\Big[ iA_L\wedge F^{(R)}\wedge F^{(R)}-iA_R\wedge
F^{(L)}\wedge F^{(L)} +\frac{i}{2}(A_L\!-\!A_R)\wedge
(F^{(L)}\wedge F^{(R)}+F^{(R)}\wedge F^{(L)})\\
&+\frac{1}{2}A_L\wedge A_L \wedge A_L \wedge F^{(L)}-
\frac{1}{2}A_R\wedge A_R \wedge A_R \wedge F^{(R)}+
\frac{i}{10} A_L\wedge A_L \wedge A_L\wedge A_L\wedge A_L\\
&-\frac{i}{10} A_R\wedge A_R \wedge A_R\wedge A_R
\wedge A_R\Big]\\
&+i\tau^3\,d\tau \wedge\Big[ (A_L\wedge A_R-A_R\wedge A_L)
\wedge (F^{(L)} +F^{(R)} ) +i A_L \wedge A_L \wedge A_L\wedge A_R\\
&-\frac{i}{2}A_L\wedge A_R\wedge A_L\wedge A_R +i A_L
\wedge A_R\wedge A_R \wedge A_R \Big]\\
&\left.+\frac{i}{20}\tau^4 (A_L-A_R)\wedge (A_L-A_R)
\wedge (A_L-A_R)\wedge (A_L-A_R)\wedge (A_L-A_R)\right\}
\end{split}
\ee
with the understanding that the contributions from the Abelian field $\Phi$ are included in the gauge fields here.
The normalization of this term is consistent with the QCD flavor anomalies~\cite{Casero:2007ae}.\

In order to extract the coupling between the solitonic components and $\Phi$ explicitly, we substituting $A_{L/R} \to \Phi dt +A_{L/R}$ in~\eqref{Omega5}  and collect the coupling terms. Recall however that $\Omega_5$ is well defined only up to total derivatives. As it turns out, it is convenient to first modify the definition of $\Omega_5$ by adding the following total derivative terms
\begin{align}
 12\widetilde \Omega_5 =12\Omega_5 &+ i\, \mathbb{T}r\,d\big[\, e^{-\tau ^2} \Phi dt \wedge (4 A_L\wedge F^{(L)}+i A_L\wedge A_L\wedge A_L-4 A_R\wedge F^{(R)}&\nonumber\\
 &\ -i A_R\wedge A_R\wedge A_R)\big] +  i \,\mathbb{T}r\, d\big[\, e^{-\tau ^2} \tau ^2 \Phi dt\wedge (-2 A_L\wedge F^{(L)}-6 A_L\wedge F^{(R)}&\nonumber\\
 &\ +i A_L\wedge A_L\wedge A_L+6 A_R\wedge F^{(L)} +2 A_R\wedge F^{(R)}-i A_R\wedge A_R\wedge A_R)\big] \,. &
\end{align}
Then we find that
\be
 \widetilde \Omega_5 =  \Omega_5\big|_{\Phi=0} +  \frac{1}{6}\Phi dt \wedge H_4^{(\Phi)} \,,
\ee  
where
\begin{align} \label{H4def}
  e^{\tau ^2}H_4^{(\Phi)} &=\mathbb{T}r\,\big[ -3 i F^{(L)}\wedge F^{(L)}+3 i F^{(R)}\wedge F^{(R)}+6 i \tau  d\tau \wedge (A_L-A_R)\wedge (F^{(L)}+F^{(R)})&\nonumber\\
  &\phantom{=}\ +3 \tau ^2 (A_L-A_R)\wedge (A_L-A_R)\wedge (F^{(L)}-F^{(R)}) &\nonumber\\
  &\phantom{=}\ + \tau ^3 d\tau \wedge (-4 i A_L\wedge F^{(R)}+4 i A_R\wedge F^{(L)}+2 A_R\wedge A_L\wedge A_L&\nonumber\\
  &\phantom{===}  -2 A_R\wedge A_R\wedge A_L-2 A_L\wedge A_L\wedge A_L+2 A_R\wedge A_R\wedge A_R)\big] \,.
\end{align}
and we have used the fact that $d\tau\wedge d\Phi =0$ since both fields are assumed to depend on $r$ only.
We note that $H_4^{(\Phi)}$ is closed, $dH_4^{(\Phi)}=0$, and exact:
\begin{align} \label{H4diff}
  H_4^{(\Phi)} &=\mathbb{T}r\, d\Big[e^{-\tau ^2} \big(-3 i A_L\wedge F^{(L)}+3 i A_R\wedge F^{(R)}+A_L\wedge A_L\wedge A_L-A_R\wedge A_R\wedge A_R &\nonumber\\
  &\phantom{=}\ +\tau ^2 (A_L-A_R)\wedge (A_L-A_R)\wedge (A_L-A_R)+3 i \tau  d\tau\wedge (A_L\wedge A_R-A_R\wedge A_L)\nonumber\\
   &\phantom{=}\ -2 i \tau ^3 d\tau\wedge (A_L\wedge A_R-A_R\wedge A_L)\big)\Big] \,. &
\end{align}

The total charge density is defined as
\be
 \varrho = -\left.\frac{\delta S_\mathrm{V-QCD}}{\delta \Phi'}\right|_\mathrm{bdry} = \int dr\, \frac{\delta S_\mathrm{V-QCD}}{\delta \Phi}\, ,
\ee
where we used the $\Phi$ equation of motion. Therefore the baryon charge is given by the coupling to $\Phi$ in the CS action: 
\be
N_c N_b = \int dr d^3x\  \frac{\delta S_\mathrm{CS}}{\delta \Phi}= \frac{iN_c}{24\pi^2} \int \ H_4^{(\Phi)} \ ,
\ee
where $N_b$ is the total baryon number. We will compute this explicitly below within the approaches considered in this article.

\section{Baryons as a thin layer of noninteracting bulk matter}  \label{sec:pointlike}

The first approach we consider is to include baryons as a layer of noninteracting solitons. The layer is located in equilibrium at a finite nonzero value of the bulk and assumed to have a zero width in the holographic direction. This setup is similar to the approach in the WSS model where the baryons were treated as point-like sources in the limit of large coupling~\cite{Bergman:2007wp,Rozali:2007rx}. When comparing to the WSS model it is useful to recall that the dynamics of chiral symmetry breaking can be discussed in terms of tachyon condensation as we did for V-QCD in Sec.~\ref{sec:VQCD}~\cite{Bergman:2007pm,Dhar:2007bz,Dhar:2008um,Jokela:2009tk}.  Notice however that in our model there will not be a limit (similar to the large coupling limit in the WSS model) in which the sizes of the solitons are suppressed. Since our approach in this section requires the extent of the baryons to be zero in the holographic direction, it should be considered as a rough approximation. Notice however that as we are neglecting the interactions between the solitons and our background solution is independent of the spacetime coordinates, the sizes of the solitons in spatial directions are irrelevant. The easiest approach is to consider the solitons to be of zero size in this direction also. We will consider another approach in Sec.~\ref{sec:homogeneous} which will take the effects due to the finite size and interactions into account at least partially.

\subsection{Setup}

In order to establish the thermodynamics in the setup, we need to compute the mass of a single soliton (integral of the expanded DBI action). As discussed above, we will essentially treat the soliton as point-like. We first consider the simplest approach, where the tachyon field $\tau$ is completely ignored -- this is a good approximation if the soliton is located very close to the boundary. We will argue how the tachyon effects can be taken into account later.
In this approximation, we obtain
\begin{align}
 S_\mathrm{DBI}^{(1)} &=  \frac{M^3 N_c}{8}  \int d^5x\, V_{f0}(\l) 
 w(\l)^2\sqrt{-\det g}\sqrt{\Xi} \left( \tilde g^{-1}\right)_s^{MN}\left( \tilde g^{-1}\right)_s^{PQ}\,&\nonumber\\
 &\qquad \times\mathbb{T}r\left(F^{(L)}_{NP}F^{(L)}_{QM}+F^{(R)}_{NP}F^{(R)}_{QM}\right) \ ,&\nonumber\\
 S_\mathrm{CS} &= \frac{N_c}{8\pi^2} \int  
 \Phi dt \wedge\mathbb{T}r\,\left[ F^{(L)}\wedge F^{(L)}- F^{(R)}\wedge F^{(R)}\right] \,.
\end{align}
Notice that the DBI action still involves the nontrivial effective metric $\left( \tilde g^{-1}\right)_s$. In order to simplify the analysis, we can rescale the coordinates and the gauge fields. Since the soliton is localized in $r$ and spatial coordinates we may rescale them but not the time. 
We define
\begin{align}
 x^i &= \sqrt{\Xi} \hat x^i \,,\quad A_{L/R}^i = \frac{1}{\sqrt{\Xi}} \hat A_{L/R}^i\,,\quad t = 
 \hat t\,,\quad A_{L/R}^t =\sqrt{f} \sqrt{1+e^{-2A}f\kappa(\l)(\tau')^2} \hat A_{L/R}^t \,,& \nonumber\\
 r &= \sqrt{f} \hat r\,,\quad A_{L/R}^r = \frac{1}{\sqrt{f}}\hat A_{L/R}^r \,.&
\end{align}
These rescalings were chosen such that the CS term remains invariant and the factors of $\left( \tilde g^{-1}\right)_s$ can be absorbed in the determinant of the rescaled metric:
\begin{align} \label{naivesoliton}
 S_\mathrm{DBI}^{(1)} &= - \frac{M^3 N_c}{8}  \int d^5\hat x\, V_{f0}(\l) 
 w(\l)^2\sqrt{f}\sqrt{-\det \hat g}\,\,\mathbb{T}r\left[\widetilde F^{(L)}_{MN}\widetilde F^{(L)MN}+\widetilde F^{(R)}_{MN}\widetilde F^{(R)MN}\right]\,, &\nonumber\\
 S_\mathrm{CS} &= \frac{N_c}{8\pi^2} \int 
 \Phi d t \wedge\mathbb{T}r\,\left[ \widetilde F^{(L)}\wedge \widetilde F^{(L)}- \widetilde F^{(R)}\wedge\widetilde F^{(R)}\right]
\end{align}
where the metric $\hat g$ is conformally flat,
\be
d s^2=e^{2 A(r)} (-d\hat t^2+d\mathbf{\hat  x}^2+d\hat r^2)\,.
\label{bamehat}
\ee
The result is similar in form to what has been found in probe brane models. Because the soliton is assumed to be localized in the $r$-direction, the result boils down to the Yang-Mills action in flat space where the solution (the BPST instanton) is known. The action may be evaluated as
\be \label{baryonmass}
 S_\mathrm{DBI}^{(1)} = - 2 M^3 N_c\pi^2  \int dt\, V_{f0}(\l) w(\l)^2\sqrt{f}e^A
 \Big|_{r=r_b} \,,\qquad S_\mathrm{CS} = N_c \int dt\, \Phi(r_b) \,,
\ee
where we reinstated the unrescaled coordinate $r$. The location of the baryon $r_b$ will be determined by minimizing the action as we will show below. For a soliton corresponding to an antibaryon the sign of the CS term is opposite. 

There is however no obvious reason (as we shall demonstrate below) why the soliton should stabilize very close to the UV boundary in our model. 
Therefore the tachyon dependence should not be discarded. We will discuss how they affect the computation starting with the CS term.

In the CS action the tachyon dependent terms are given in~\eqref{H4def} and~\eqref{H4diff}. For a soliton localized in the $r$-direction the CS term can be written as
\be \label{SCSinst}
 S_\mathrm{CS} \simeq \frac{iN_c}{24\pi^2} \Phi(r_b) \int  dt \wedge  H_4^{(\Phi)} \ .
\ee
We expect that the integral here  is quantized in units of $24 \pi^2 i$ also when the soliton is not close to the boundary, and the result therefore is the same as in~\eqref{baryonmass}. Since the integral couples simply to $\Phi(r_b)$ this is consistent with the baryons carrying a fixed charge. Indeed as the form $H_4^{(\Phi)}$ is exact, the integral becomes a boundary term, which suggests that the quantization can be read off by inserting the asymptotic form of the soliton solution in this expression. However, in the absence of an explicit soliton solution which would take into account the coupling to the tachyon, we have not been able to prove this.

This result implies in particular that the integrals over the various terms in~\eqref{H4def} on the soliton solution will need to grow large in order to compensate the factor $e^{-\tau^2}$ in the expression of this form if the soliton is located deep in the IR. That is, the amplitude and/or size of the soliton needs to grow large. 
Therefore, it is essential that for the approximations done in this section to work, the baryon is not located very deep in the IR.

Without better control of the soliton solution, it is hard to evaluate its contribution to the DBI action, i.e., the soliton mass, in the presence of the tachyon corrections. The main addition due to the tachyon is the factor $e^{-\tau^2}$ in the potential of the DBI term, see~\eqref{SDBI1final}. The quantization argument of the CS term suggests that the contribution of the soliton grows if it is moved towards the IR such that it roughly cancels this term. Therefore our best guess for the effects of the tachyon is that they are absent at least when the soliton is not too deep in the IR, that is, we will also use the expressions in~\eqref{baryonmass} in the presence of the coupling to the tachyon. We remind that we will consider a different approach in Sec.~\ref{sec:homogeneous} which will capture the coupling to the tachyon.

In the WSS model the baryon action is obtained through a D4 action, with the brane wrapping the $S^4$ of the geometry, or equivalently by considering an expansion of the D8 actions at small gauge fields~\cite{Sakai:2004cn,Rozali:2007rx}. Doing a simple minded mapping of this approach to our model,  the solitonic solutions should correspond roughly to adding a D0 brane in the configuration. Indeed, noticing that $\sqrt{f}e^A=\sqrt{-g_{tt}}$, the first term in~\eqref{baryonmass} takes the form of an action for a D0 brane sitting at the location of the baryon (with a certain $\l$ dependent potential). 

The final action for a baryon gas with constant density is then obtained by a ``convolution'' which amounts to integrating the above actions $\int d^3x\, n_b$ to the actions in~\eqref{baryonmass}:
\be \label{baryonmass2} 
 S_\mathrm{DBI}^{(1)} = - 2 M^3 N_c\pi^2  \int d^4x\, n_b\, V_{f0}(\l) w(\l)^2\sqrt{f}e^A
 \Big|_{r=r_b} \,,\qquad S_\mathrm{CS} = N_c \int d^4x\, n_b\, \Phi(r_b) \,.
\ee

\subsection{Equations of motion and boundary conditions}\label{sec:pointlikeeom}

The complete action of the model is given by the sum of the terms in~\eqref{glueaction},~\eqref{SDBI0}, and~\eqref{baryonmass2} in the current approach.
We define the bulk charge density as
\be \label{rhodef}
 \rho=-\frac{\delta  S_\mathrm{DBI}^{(0)}}{\delta \Phi'} =  -\frac{M^3 N_c N_f V_{f0}(\l)e^{-\tau^2}e^{A}w(\l)^2\Phi'}{\sqrt{1+ e^{-2A}f\kappa(\l)(\tau')^2-e^{-4A}w(\l)^2(\Phi')^2}}\ .
\ee
The equation of motion for $\Phi$ implies
\be
  \rho'(r) = -N_c n_b \delta(r-r_b) \,. 
\ee
The thermal gas solutions extend form $r=0$ (UV boundary) to $r=\infty$ (IR singularity)\footnote{Actually, for the choice of action for which we carry out numerical analysis below, the TG solutions turn out to be subdominant.}. 
For these configurations all the charge originates from the baryons, and therefore $\rho(r) =0$ for $r>r_b$. Consequently, $\rho(r) = N_c n_b \equiv \rho_b$ for $0<r<r_b$. The black hole solutions extend from the boundary ($r=0$) to a horizon at some value $r=r_h$ of the bulk coordinate. For them, part of the charge may be hidden behind the horizon. Then $\rho$ is nonzero everywhere, and constant except for the discontinuity at the baryon location: $\rho(r) = \rho_h$ for $r_b<r<r_h$ and $\rho(r) = \rho_h + \rho_b \equiv \varrho$ for $0<r<r_b$.

The gauge field is obtained by inverting~\eqref{rhodef},
\be \label{Phidersol}
 \Phi' = - \frac{\hat \rho}{V_{f0}(\l)e^{-\tau^2}w(\l)^2e^A}\frac{G}{\sqrt{1+K}} \ ,
\ee
where
\be
 G = \sqrt{1+e^{-2A}f\kappa(\l)(\tau')^2} \ , \qquad K = \frac{\hat \rho^2}{\left(e^{3A}V_f(\lambda ,\tau )w(\l)\right)^2} \ ,
\ee
and integrating over $r$. Here we defined the normalized density as $\rho= M^3 N_c N_f \hat\rho$. We will discuss below how the constant of integration is fixed.

When $r \ne r_b$, the other equations of motion are
\begin{align}
\label{taueq}
\frac{d}{dr}\left[ \frac{e^{3 A} f\sqrt{1+K} \kappa (\lambda)  V_f(\lambda,\tau)\tau '}{G}\right] = \frac{e^{5 A} G \frac{\partial}{\partial \tau}V_f(\lambda,\tau)}{\sqrt{1+K}}\,,\\
\label{ceq}
12 f A'^2 +3 f'A'-\frac{4  f \lambda '^2}{3 \lambda ^2}-e^{2 A}V(\lambda )+ \frac{x_f e^{2 A}\sqrt{1+K} V_f(\lambda ,\tau )}{G} = 0\,, \\
6 f A''+6 f A'^2+3 f'A'+\frac{4 f\lambda '^2}{3 \lambda ^2}-e^{2 A}V(\lambda )+x_f e^{2 A}G \sqrt{1+K} V_f(\lambda ,\tau )=0\,, \\
f'' +3 f'A' - x_f e^{2 A}\frac{G K}{\sqrt{1+K}} V_f(\lambda ,\tau )=0 \,,
\end{align}
where
 we already inserted the solution for $\Phi'$ from~\eqref{Phidersol}. For the thermal gas solutions, $f=\mathrm{const.}$ for $r>r_b$. At $r=r_b$ we will also need the junction conditions for the various functions which are derived in Appendix~\ref{app:junctions}.

The solutions satisfy the following boundary conditions in the UV:
\be
A(r) \approx -\log r, \qquad f(r) \approx 1, \qquad \lambda(r) \approx -\frac{1}{b_0\log(r\Lambda)}, \qquad \tau(r) \approx m_qr(-\log(r\Lambda))^{-\gamma_0/b_0},
\ee
with $b_0 = \frac{1}{24 \pi^2}(11 - 2x_f)$ and $\gamma_0 = \frac{3}{16 \pi^2}$\@.
The boundary condition for $\lambda$ can here be taken as the definition of $\Lambda$\@.
Dimensionful quantities are computed relative to the energy scale defined by $\Lambda$, and putting $\Lambda$ to a physically reasonable value, one can match with QCD\@. In all numerical examples considered in this article, we set the quark mass $m_q$ to zero. 

In the IR, if a black hole is present, the solutions satisfy regularity conditions on the horizon.
The thermal gas solution, which has no black hole, is obtained by taking a black hole solution and letting the horizon area approach zero.
This is in accordance with the requirement that the IR singularity should be of the ``good'' type, i.e.~it should be possible to clock the singularity with a horizon of infinitesimal area.

\subsection{Location of the baryon}\label{sec:pointlikelocation}

In order to determine the location of the baryon, because each baryon carries a fixed charge, we need to study the system in canonical ensemble~\cite{Bergman:2007wp,Rozali:2007rx}.
The Legendre transformed zeroth order action reads
\begin{align}
 \widetilde S_\mathrm{DBI}^{(0)} &= S_\mathrm{DBI}^{(0)} -\int d^4x\,\Phi(0)\rho+S_\mathrm{CS} = S_\mathrm{DBI}^{(0)} + \int d^5x\, \rho \Phi' &\nonumber\\
 &= - M^3 N_c N_f \int  d^5x\,V_{f0}(\l)e^{-\tau^2}e^{5A}\sqrt{1+e^{-2A}f\kappa(\l)(\tau')^2}\sqrt{1+\left(\frac{\hat \rho}{e^{3A}w(\l)V_{f0}(\l)e^{-\tau^2}}\right)^2} &\nonumber\\
 &= - M^3 N_c N_f \int  d^5x\,V_{f0}(\l)e^{-\tau^2}e^{5A}G\sqrt{1+K}\,,
\end{align}
where the integrand has a discontinuity at $r=r_b$ and we also included the CS contribution. In order to determine $r_b$, we need to minimize the full action 
\begin{align} \label{LTaction}
 \widetilde S_\mathrm{DBI} = \widetilde S_\mathrm{DBI}^{(0)} + S_\mathrm{DBI}^{(1)}&=- M^3 N_c N_f \int  d^5x\,V_{f0}(\l)e^{-\tau^2}e^{5A}G\sqrt{1+K}&\nonumber\\
 &\phantom{=}\ - 2 M^3 N_c\pi^2  \int d^4x\, V_{f0}(\l) w(\l)^2\sqrt{f}e^A n_b
 \Big|_{r=r_b}
\end{align}
varying $r_b$. As we keep all sources fixed it is enough to evaluate the derivative with respect to the explicit dependence on $r_b$. After the Legendre transformation, this appears in the source term and through the discontinuity of $\hat \rho$. Taking the derivative gives the condition (see Appendix~\ref{app:junctions})
\begin{align} \label{baryoncond}
&\phantom{=} V_{f0}(\l) e^{-\tau^2}e^{5A}\frac{1}{G(r_b^+)}\left[\sqrt{1+K(r_b^+)+G(r_b^+)^2 (K(r_b^-)-K(r_b^+))}-\sqrt{1+K(r_b^+)}\right]_{r=r_b}&\nonumber\\
&=-2\pi^2 M^3\hat \rho_b  \sqrt{f}e^A\left\{  \langle\l' \rangle \frac{d}{d\l}\left[V_{f0}(\l) w(\l)^2
\right] +  V_{f0}(\l) w(\l)^2 \left[\frac{\langle f'\rangle}{2 f} + \langle A'\rangle\right]\right\}_{r=r_b} \,,&
\end{align}
where $g(r_b^\pm) \equiv \lim_{\epsilon \to 0 +}\, g(r\pm \epsilon)$ and the averaged derivatives are defined by $\langle g'\rangle \equiv (g'(r^+_b)+g'(r^-_b))/2 = \lim_{\epsilon \to 0 +} (g'(r_b+\epsilon)+g'(r_b-\epsilon))/2$. 

In the limit $\hat \rho \to 0$ the first term in~\eqref{baryoncond} vanishes as $\propto \hat\rho^2$, $f$ tends to one, and the derivatives become continuous. Therefore, the condition becomes 
\be \label{masscond}
\frac{d}{dr}\left[V_{f0}(\l) w(\l)^2e^A \right]_{r=r_b} = 0 \ ,\qquad \left(\hat \rho \to 0\right) \,.
\ee
That is, the term defining the soliton mass in~\eqref{baryonmass} should have a minimum in the bulk, and the minimum should be quite close to the boundary: otherwise the baryon will fall deep in the IR in the regime where the coupling of the soliton to the tachyon field becomes important. Our approximation, where the tachyon is essentially neglected will fail in this region.

In order to realize the minimum of the baryon mass at finite $r$  we need to choose the potential $w(\l)$ differently from earlier literature (see~\cite{Arean:2013tja,Arean:2016hcs,Jokela:2018ers}). The easiest way\footnote{In principle one may find nontrivial solutions to~\eqref{masscond} without this property, but numerical studies show that one needs to introduce potentials (e.g. $w(\l)$) with peculiar structures, which are likely to cause other issues.} to guarantee a minimum is to require that the combination in the square brackets of~\eqref{masscond} diverges in the IR (because it does diverge in the UV). If $V_{f0}\sim \l^{v_p}$ and $w \sim \l^{-w_p}$, this means that
\be \label{wpmassconstraint}
 w_p\le \frac{v_p}{2} - \frac{1}{3} \ .
\ee
Because we need to choose $v_p \le 10/3$~\cite{Arean:2013tja} we have that $w_p \le 4/3$. In practice, however, the bound is tighter: we have not been able to construct numerically regular backgrounds for potentials with $v_p \gtrsim 3$. Here we use $v_p=2$, so that $w_p \le 2/3$.
We will present some results from the numerical analysis of the model for such potentials with $w_p=2/3$ below. 

However, as we discussed in Sec.~\ref{sec:backgroundthermo} and in Appendix~\ref{app:wasympt}, the phase diagram for this choice of $w_p$ has an undesired structure even in the absence of baryons: the thermal gas phase, which is identified as the confined chirally broken phase in QCD, becomes subdominant and is replaced by a phase with ``tiny'' black hole solutions. 
The requirement of the TG phase to be dominant leads to $w_p \ge 4/3$ which is in contradiction with~\eqref{wpmassconstraint} (inserting $v_p \lesssim 3$ which is in turn required for regular backgrounds). 

This is, however, a problem with the approximations done in this section rather than a problem in the model: in Sec.~\ref{sec:homogeneous} we will demonstrate that the coupling of the soliton to the tachyon (which is basically not included in the thin layer approximation scheme) will prevent the soliton from falling in the IR. Therefore, after the coupling to the tachyon has been added, it is actually possible to choose $w(\l)$ in the same way as in earlier literature and as required by the analysis of Appendix~\ref{app:wasympt}. Then a physically reasonable phase diagram with both baryons and a TG phase can be obtained.

\subsection{Thermodynamics}\label{sec:pointlikethermo}

The grand potential is given by
\be \label{Fdef}
 \Omega = -\left[S_\mathrm{glue}+S_\mathrm{DBI}^{(0)}+S_\mathrm{GH}+S_\mathrm{source}\right]_\mathrm{on-shell} \ ,
\ee
where the source term is
\be
 S_\mathrm{source} = - 2 M^3 N_c N_f\pi^2  \int d^4x\, V_{f0}(\l) w(\l)^2\sqrt{f}e^A M^3 \hat \rho_b\Big|_{r=r_b}  + M^3 N_cN_f \int d^4x\, \Phi \hat \rho_b \Big|_{r=r_b} \ ,
\ee
the Gibbons-Hawking term $S_\mathrm{GH}$ is given in Appendix~\ref{app:junctions}, and the minus sign in~\eqref{Fdef} appears because we wrote our actions in Lorentzian signature.

In order to establish the thermodynamics of the system and check its consistency, we need to determine the integration constant in the definition of $\mu$. We reproduce here the basic arguments and delegate details to Appendix~\ref{app:junctions}.

As it turns out, if we choose a gauge where $\mu =\Phi(0)$, we need to require that the source term vanishes on-shell by setting
\be \label{Phirbval}
 \Phi(r_b) = 2 M^3 \pi^2  V_{f0}(\l) w(\l)^2\sqrt{f}e^A \Big|_{r=r_b} \equiv \mu_c \ .
\ee
This fixes the integration constant and ensures that the variation of the free energy follows the first law: if the source vanishes, its variation vanishes as well, and then the variation of the bulk term gives the first law by the standard calculation. 

For black hole solutions,~\eqref{Phirbval} can also be derived as the equilibrium condition for moving charge from the baryons to behind the horizon. In order to see this, consider the variation of the Legendre transformed action~\eqref{LTaction} with respect to the charge behind the horizon $\hat \rho_h$ such that the total charge $\hat \varrho = \hat \rho_h+\hat\rho_b$ is fixed. This gives
\begin{align}
 \frac{1}{M^3N_cN_f}\frac{\delta \widetilde S_\mathrm{DBI}}{\delta \hat\rho_h} &= - \int d^4x \int_{r_b}^{r_h}dr\ \frac{V_f e^{5A} G}{\sqrt{1+K}}\frac{K}{\hat\rho_h} + 2 M^3 \pi^2 \int d^4x\ V_{f0} w^2 \sqrt{f} e^A\Big|_{r=r_b} &\nonumber\\
 &= V_4\left[-\Phi(r) + 2 M^3 \pi^2  V_{f0} w^2 \sqrt{f} e^A\right]_{r=r_b}
\end{align}
where we used  the EoM~\eqref{rhodef} on the second line. Therefore, requiring the variation to vanish results in~\eqref{Phirbval}.

Naively one might think that the discontinuities in the bulk profile induced by the point like source may also lead to nonzero terms at $r=r_b$ which could violate the first law. However, by replacing the delta distribution of the source by a smooth approximation, we see that no such terms can arise: the variation of the Lagrangian is still a total derivative and only boundary terms at $r=0$ are generated. In the point-like limit the junction conditions for the bulk fields at $r=r_b$, which are given above, guarantee that all contributions from $r=r_b$ in the variation of the grand potential cancel. We show this explicitly in Appendix~\ref{app:junctions}. 

By using~\eqref{rhodef}, for the thermal gas solution the chemical potential is then given by
\be
 \mu = \Phi(r_b) - \int_0^{r_b} dr\, \Phi'(r) = \mu_c + \int_0^{r_b} dr\, \frac{\hat \rho}{V_{f0}(\l)e^{-\tau^2}w(\l)^2e^A}\frac{G}{\sqrt{1+K}} \ .
\ee
For black hole solutions,~\eqref{Phirbval} sets another constraint which will fix the ratio of the charge of the baryons to the total charge. For the black holes we simply have the  formula
\be
 \mu =\int_0^{r_h} dr\, \frac{\hat \rho}{V_{f0}(\l)e^{-\tau^2}w(\l)^2e^A}\frac{G}{\sqrt{1+K}} \ . 
\ee

The free energy in each phase can then be obtained by integrating
\be
 d \Omega = - \varrho d \mu - s dT
\ee
where $\varrho$ is the boundary value of the charge density, $\varrho = \rho(0)$, for the TG solutions the entropy vanishes, and for the BH solutions the entropy and the temperature are given by standard BH thermodynamics (see also~\cite{Alho:2012mh,Alho:2013hsa}).

\subsection{Phase diagram}
Using the results of the previous sections, we can compute the phase diagram.
The non-baryonic solutions are constructed as in \cite{Alho:2012mh,Alho:2015zua,Alho:2013hsa}\@.
The baryonic solutions are constructed in the same way, i.e.~by shooting from the horizon, with the difference that when it is encountered that equation \eqref{baryoncond} is satisfied, a discontinuity is inserted in accordance with the junction conditions derived in Appendix \ref{app:jcs}\@. We choose $x_f=1$ and only consider solutions with vanishing quark mass.
Both the non-baryonic and the baryonic solutions are then used to integrate the first law as described in Sec.~\ref{sec:pointlikethermo}\@.

This procedure allows one to compute the phase diagram given a set of potentials.
The potentials which are used in the following are given in Appendix \ref{app:potentials}\@.
They have been obtained by fitting to lattice data in the vicinity of $\mu = 0$ \cite{Jokela:2018ers}\@.
Note that in the thin layer approximation, we need to satisfy the bound~\eqref{wpmassconstraint} in order to have stable baryon matter. Consequently we choose a $w$-function with the asymptotics 
$w \sim \lambda^{-2/3}$ in the IR\@. The explicit choice is given 
in~\eqref{eq:wpotpointlike}\@.\footnote{Note that in Appendix \ref{app:potentials}, there are two sets of potentials 
where the most significant difference between the two choices  for $w(\lambda)$ in \eqref{eq:wpotpointlike} and in \eqref{eq:wpothomogeneous} is given by the IR asymptotics. We will use the second set later on.}
Also note that the $w$-potential was, by tuning the stabilization point, used to set the baryon mass (given by $-S_\mathrm{DBI}^{(1)}$ in~\eqref{baryonmass}) to roughly the correct value.

As we pointed out in Sec.~\ref{sec:backgroundthermo} and in Sec.~\ref{sec:pointlikelocation}, the choice for \eqref{eq:wpotpointlike} also has an unintended consequence.
Denoting $w \sim \lambda^{-w_p}$ for the power $w_p$ of the $w$-potential in the IR, one can show that the thermal gas phase is always subdominant  at nonzero values of the chemical potential unless $w_p \geq 4/3$\@. This means that a limitation of this approximation is that it is incompatible with having a thermal gas phase at small but nonzero $T$ and $\mu$\@.\footnote{Note that exactly at $\mu = 0$, these potentials still display a thermal gas phase, because as $\mu \rightarrow 0$, the size of the black hole shrinks to zero and the geometry approaches that of the thermal gas solution.}
The reason for this is discussed in more detail in Appendix \ref{app:wasympt}\@.
We will use the $w$-potential given by \eqref{eq:wpotpointlike} despite the issue with the TG phase, because for this approximation to yield a non-trivial result, we need to have stable baryons.

Moreover this choice of $w$-function destroys the Silver Blaze property of QCD 
(see~\cite{Armoni:2014wla}): the pressure is no longer independent of the chemical potential at zero temperature and at nonzero chemical potentials up to a critical value. This happens because the thermal gas depends nontrivially on the chemical potential.
It also means that, because the thermal gas phase is replaced by a small black hole, confinement properties are altered.
In holography one can investigate confinement properties from the behavior of a string suspended from two points on the boundary.
In particular, one can calculate the Wilson loop by computing the on-shell action of such a string~\cite{Rey:1998ik,Maldacena:1998im,Kinar:1998vq,Bak:2007fk}. 
In this work, we define a phase as confining if the Wilson loop one computes in this way has a branch exhibiting an area law.
The crucial difference between the confinement properties of a thermal gas and the confinement properties of the solutions one obtains in this section is that even if the Wilson loop has such a branch, these can correspond to string solutions which are unstable and subdominant to disconnected strings extending to the black hole from the boundary.
Note that this instability is a separate effect from the ``usual'' breaking of the string in QCD due to a pair creation of light quarks. This pair creation effect is not included in the classical string computation we discuss here. Therefore the instability which we observe here is indeed undesirable for QCD. 
For a more detailed discussion of confinement in a similar geometry, we refer to \cite{Gursoy:2018ydr}\@.
Note that thermodynamic properties like the equation of state are much less affected by this change in geometry, as they do not probe the deep IR\@.

\begin{figure}[t]
\centering
\includegraphics[width=\textwidth]{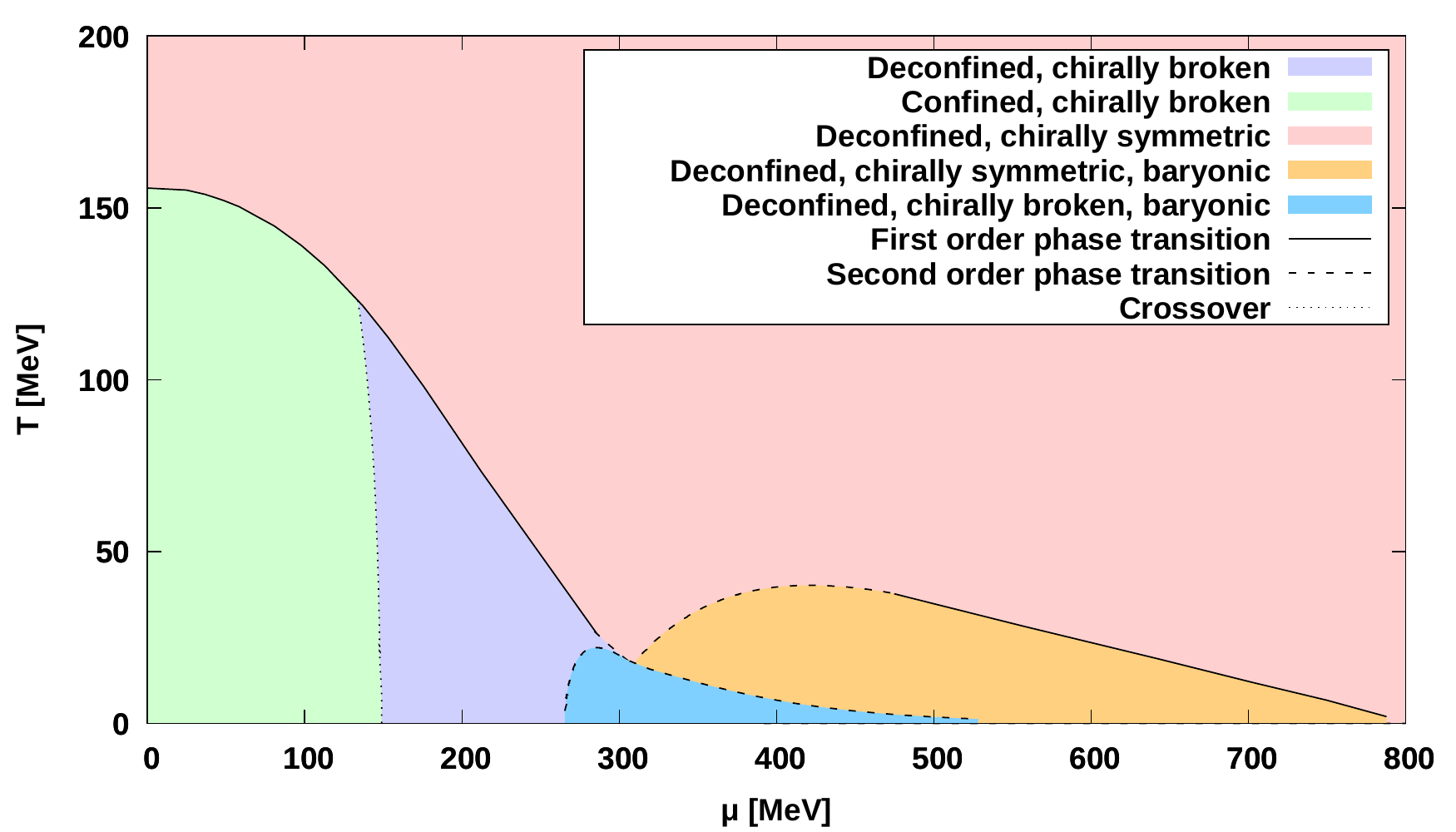}
\caption{\label{fig:phases}The phase diagram for the thin layer approximation in the $(\mu,T)$-plane. It can be seen that we have two phases in which baryon number charge is located outside the black hole. Note that due to numerical accuracy the phase transitions cannot be accurately continued to the $\mu$-axis, but they do all reach it.}
\end{figure}

There are three order parameters by which we label the phases in the phase diagram:
\begin{itemize}
\item Chiral symmetry: If the chiral condensate $\langle\bar qq\rangle$ is zero, a phase displays chiral symmetry.
If it is nonzero, chiral symmetry is broken.
\item Appearance of baryons: If the $U(1)$ charge associated to quark number\footnote{Note that quark number is equal to baryon number divided by the number of colors.} is all located behind a black hole horizon, that phase does not have baryons.
If, however, a part of the baryon number charge is located at some point in the bulk, we say that a phase is baryonic.
Note that even with some of the charge located outside the black hole, part of the charge always remains behind the black hole, making the baryonic phases a sort of mixture of baryons and quark matter.
\item Confinement: A phase is called confining if the Wilson loop exhibits an area law.
As was discussed before, this does not imply confinement in the usual sense.
Instead, more weakly, it implies that the properties one usually associate with confinement, like a mass gap and the inability to pull apart two quarks, are only approximately satisfied.
\end{itemize}
By comparing these properties, there are in principle 8 phases we can have in the phase diagram.
Of these, 5 are realized by the model. 
These are the 4 possibilities for deconfined phases, plus a confined, chirally broken nonbaryonic phase.
Comparing free energies of these 5 phases, one obtains the phase diagram in Fig.~\ref{fig:phases}\@.
Note that as one expects at $\mu = 0$, there is a first order phase transition between a chirally symmetric, deconfined QGP phase, and a chirally broken confined phase.
As $\mu$ is increased at fixed $T$, the chirally broken phase becomes deconfined as well.
Since the geometry changes smoothly as confinement is lost, the transition is a crossover.
Also note that the first order transition extending from $\mu = 0$ ends in a critical point, turning into a second order phase transition just before the baryonic phases appear.
Another critical point can be found between the chirally symmetric deconfined QGP phase and the chirally symmetric deconfined baryonic phase.

Now let us discuss the baryonic phases.
The most important observation is their location in the phase diagram.
In particular, note that the chirally broken baryonic phase appears at roughly $\mu=270\,\text{MeV}$\@.
This number should be roughly equal to the baryon mass over the number of flavors, which for QCD is (up to the binding energy of nuclear matter) $m_\text{proton}/3 \approx 313\,\text{MeV}$\@.
This, while not quantitatively the same, is qualitatively in the right range.
In particular, one can note that while the potentials were chosen to reproduce the correct baryon mass, the fact that the location of the transition indeed appears in the appropriate location is a non-trivial observation.
Another thing to note is that the baryonic phases disappear above a finite value of $\mu$\@.\footnote{The first order transition between the chirally symmetric baryonic phase and the chirally symmetric nonbaryonic phase has a large numerical uncertainty. However, the chirally symmetric baryonic solutions stop existing altogether at a finite value of $\mu$, so even if we cannot pinpoint to great accuracy where the chirally symmetric baryonic phase becomes thermodynamically disfavored, we can say with certainty that this will happen at some finite value of $\mu$.}
A last observation is that the properties of the chirally symmetric baryonic phase seem somewhat contradictory.
On the one hand, chiral symmetry is exact, meaning that there is no mechanism by which the quarks can gain mass.
On the other hand, these massless quarks form bound states in the form of baryons.
This phase could perhaps be studied in future work by studying the excitations of the theory in that region of the phase diagram.

It is clear that this approximation has its shortcomings, the most serious of which is that from QCD we expect the confined solutions to be described by a thermal gas, while in this approximation we obtain a phase which is not confining in the usual sense.
In the next section, we take a different approach, in which these problems are not present.

\section{Baryons from a homogeneous bulk gauge field} \label{sec:homogeneous}

The approximation considered in the previous section obviously has some shortcomings, as we already pointed out. There is no reason to expect the baryonic soliton to be small in V-QCD, unlike in the WSS model where the size of the soliton $\sim 1/\sqrt{\lambda}$ goes to zero in the limit of strong coupling. Moreover, we basically neglected the tachyon, forcing us to choose a specific kind of potentials which keep the soliton close to the UV boundary where the tachyon is small. This choice of potentials was seen to cause problems with confinement properties. Furthermore, as it turns out, such potentials are slightly disfavored by the fit to lattice data~\cite{Jokela:2018ers}. When the baryon is no longer close to the boundary, the amplitude and/or size of the soliton must actually grow in order to account for the suppression due to the exponential factor $e^{-\tau^2}$ in order to keep the baryon charge fixed (see~\eqref{SCSinst} and~\eqref{H4def}).

If the soliton becomes sizable in the IR, configurations with a high density of solitons, dual to dense baryonic matter in QCD, may be described better by a homogeneous non-Abelian gauge field configuration than separate solitons. This is what we will attempt here. Interestingly, as it turns out, the approach will closely resemble the approximation with a thin noninteracting layer carried out above even though the starting point is completely different.
A similar approach has been suggested in the context of the WSS model in~\cite{Rozali:2007rx} and further developed in~\cite{Li:2015uea} (see also~\cite{Elliot-Ripley:2016uwb} for a slightly different setup). 
We will treat the baryons as probe on top of the TG background in this section. 
This makes sense as the DBI action discussed above is known precisely only to leading nontrivial order in the non-Abelian field strengths of the solitons.

The basic idea of the setup is as follows. We consider a system with a high density of baryons (comparable to the saturation density on the QCD side) on top of the TG background and divide the space in to three regions in the $r$-direction:
\begin{enumerate}
 \item Region close to the boundary, $r\ll r_c$, where $r_c$ is roughly the location of the soliton ``centers''. At high density, the configuration in this region is assumed to be well approximated by a homogeneous baryon field.
 \item The region in the middle, $r \sim r_c$. In this region, the configuration is highly inhomogeneous and nontrivial.
 \item The region in the IR, $r \gg r_c$. In this region the baryon field is again taken to be homogeneous.   
\end{enumerate}
The idea is then that when the baryon density is high, the second, inhomogeneous region is not important for the main features of the phase diagram, and may be ignored. Its effect on the solutions is modeled through a discontinuity of the baryon field, as we shall discuss below. This is an uncontrolled approximation, but as we shall see, the results are encouraging.

\subsection{Setup}

We will only consider $SU(2)$ solitons in the thermal gas background here. We will use the first order series approximation to the DBI action rather than the full DBI (which is not known for non-Abelian fields). Our ansatz 
\be \label{homans}
 A_L^i = -A_R^i = h(r) \sigma^i 
\ee
respects chiral symmetry and parity~\cite{Pomarol:2007kr,Pomarol:2008aa}. 
  
The ansatz~\eqref{homans} immediately leads to an issue which has also been observed in the WSS model~\cite{Rozali:2007rx}. Namely inserting it in the expression of the baryon charge gives
\be
 \int dt\wedge H_4^{(\Phi)} = 48i \int d^5 x\, \frac{d}{dr}\left[ e^{- b\, \tau(r)^2}h(r)^3(1-2b\, \tau(r)^2)\right] \ ,
\ee
where we reinstated the coefficient $b$. If $h(r)$ is a smooth function this evaluates to a boundary term. Both the UV and IR contributions however vanish: the diverging tachyon sets the action to zero in the IR, and the baryon field $h(r)$ vanishes in the UV due to boundary conditions. Therefore the baryon density is zero.

A nonzero baryon density can however arise from a discontinuity of the function $h(r)$. 
As discussed above, we use such an abrupt discontinuity of the function $h(r)$  to model the intermediate inhomogeneous regime, which then gives rise to a nonzero baryon density. 

We then work out the action for the homogeneous ansatz. The DBI term in~\eqref{SDBI1final} simplifies to 
\begin{align}
 S_\mathrm{DBI}^{(1)} = -12 M^3 N_c \int d^5x\,V_{f0}(\l) e^{-\tau^2} e^{5A} \sqrt{\Xi}\bigg[&\kappa(\l)\tau^2e^{-2A}h^2+w(\l)^2e^{-4A}h^4&\nonumber\\
& +\frac{1}{4} w(\l)^2e^{-4A} f \Xi^{-1} \left(h'\right)^2\bigg] &
\end{align}
with $\Xi$ given in~\eqref{Rdef}. Since the charge is only sourced by the baryons, which are treated as probes, we could also work with only the leading perturbation due to the charge in $\Xi$. We choose however to keep the nonlinear dependence on the charge here.

Importantly, the homogeneous ansatz satisfies the consistency condition~\eqref{conscond}. Putting this expression together with the DBI action in the absence of solitons and the CS term then gives the total action 
\begin{align} \label{Shfinal}
S_h &=  S_\mathrm{DBI}^{(0)}+S_\mathrm{DBI}^{(1)} + S_\mathrm{CS} = - 2 M^3 N_c \int d^5x\,V_{f0}(\l) e^{-\tau^2} e^{5A} \sqrt{\Xi}\bigg[1+6\kappa(\l)\tau^2e^{-2A}h^2&\nonumber\\
 &\phantom{====}+6w(\l)^2e^{-4A}h^4+\frac{3}{2} w(\l)^2e^{-4A} f \Xi^{-1} \left(h'\right)^2\bigg] - \frac{2N_c}{\pi^2} \int d^5x\, \Phi  \frac{d}{dr}\left[ e^{-b\,\tau^2}h^3(1-2b\,\tau^2)\right] &
\end{align}
As we argued above, this action should be only trusted away from the vicinity of $r=r_c$ where $h$ is discontinuous. In particular we ignore the singular contributions which arise form the derivative of the discontinuities.\footnote{For example, for the CS term, this means effectively adding the term $  S_\mathrm{Disc} = \frac{2N_c}{\pi^2} \int d^4x\, \Phi\, e^{-b\,\tau^2}(1-2b\,\tau^2)\ \mathrm{Disc}\, h^3\big|_{r=r_c} $ which cancels the singular contribution.} In general, the prescription which we will use, amounts to interpreting the integrals as
\be
 \int_0^\infty dr \mapsto \left(\int_0^{r_c^-}+\int_{r_c^+}^\infty\right)dr \equiv \lim_{\epsilon \to 0 +}\left(\int_0^{r_c-\epsilon}+\int_{r_c+\epsilon}^\infty\right)dr\ .
\ee

From the action~\eqref{Shfinal}, which depends on $\Phi'$ through $\Xi$, we derive the charge density
\be \label{rhohomdef}
\rho = -\frac{\delta S_h}{\delta \Phi'} = -\frac{V_\rho}{\sqrt{\Xi}} \bigg[1+6\kappa(\l)\tau^2e^{-2A}h^2+6w(\l)^2e^{-4A}h^4-\frac{3}{2} w(\l)^2e^{-4A} f \Xi^{-1} \left(h'\right)^2\!\bigg]w(\l)^2e^{-4A} \Phi'
\ee
where we abbreviated
\be
V_\rho =2 M^3 N_c V_{f0}(\l) e^{-\tau^2} e^{5A} \ .
\ee
The $\Phi$ equation of motion implies
\be
 \rho' = - \frac{d}{dr}\frac{\delta S_h}{\delta \Phi'} = -\frac{\delta S_h}{\delta \Phi}= \frac{2N_c}{\pi^2}  \frac{d}{dr}\left[ e^{-b\,\tau^2}h^3(1-2b\,\tau^2)\right]
\ee
except for $r=r_c$ where $h$ is discontinuous. Our prescription stipulates that there are no $\delta$-function contributions at this point, so $\rho$ is continuous. We obtain that
\be \label{rhosolhom}
\rho = \left\{\begin{array}{lr}
               \varrho  + \frac{2N_c}{\pi^2} e^{-b\,\tau^2}h^3(1-2b\,\tau^2) \, , \qquad&(r<r_c) \\
                \frac{2N_c}{\pi^2} e^{-b\,\tau^2}h^3(1-2b\,\tau^2) \, , \qquad&(r>r_c)
              \end{array}\right.
\ee
where $\varrho$ is the density at the boundary, i.e., the physical baryon density. The continuity of $\rho$ implies that it is given in terms of the discontinuity of $h$ as
\be \label{rho0def}
 \varrho = \frac{2N_c}{\pi^2} e^{-b\,\tau(r_c)^2}(1-2b\,\tau(r_c)^2)\ \mathrm{Disc}\, h^3(r_c) \ ,
\ee
where the discontinuity is defined as $\mathrm{Disc}\, g(r) \equiv \lim_{\epsilon \to 0+}\left(g(r+\epsilon)-g(r-\epsilon)\right)$.

\subsection{Location of solitons and consistency of thermodynamics}

In this subsection we discuss the minimization of the action in particular to determine the location of the discontinuity $r_c$. Before going to the precise analysis, we point out how the main features arise from the action we wrote down above. First, 
as in Sec.~\ref{sec:pointlike} we need to work at fixed baryonic charge, rather than chemical potential. Therefore, $\varrho$ in~\eqref{rho0def} is kept fixed. This means that $\mathrm{Disc}\, h^3$ must diverge if $r_c$ is taken to deep in the IR (where the tachyon diverges) and also at the point where $2b\,\tau(r_c)^2 =1$. This necessarily means that the DBI action and consequently also the free energy diverge at these values of $r_c$. In particular, the coupling to the tachyon therefore prevents the baryon from falling in the IR.

Moreover, since~\eqref{rho0def} gives roughly $\varrho \sim h^3$, we expect that the free energy behaves as $F \sim h^2 \sim \varrho^{2/3}$ at small $\varrho$ and as $F \sim h^4 \sim \varrho^{4/3}$ at larger $\varrho$. Taking the derivative with respect to $\varrho$, we obtain that $\mu \sim \varrho^{-1/3}$ at small $\varrho$ and  $\mu \sim \varrho^{1/3}$ at larger $\varrho$. This indicates that $\mu$ has a minimum and there is a first order phase transition between the empty TG phase and the baryonic phase. Moreover, in the stable phase (larger $\varrho$) the zero temperature speed of sound obeys $c_s^2 = d \log \varrho/d \log \mu \approx 1/3$, i.e., it is roughly given by the conformal value. We will confirm numerically below that the transition is of first order and that the speed of sound is close to the conformal value.

We then move on to the precise analysis. In order to study the system at fixed charge, we perform a Legendre transformation
\be
 \widetilde S_h = S_h - \int d^4x \Phi(0) \rho(0) = S_h + \int d^5x\,  \frac{d}{dr}\left[\Phi\rho\right] \  .
\ee
Notice that (using our prescription at $r=r_c$)
\be
 S_\mathrm{CS} 
 = \int d^5x\, \Phi \rho' \ . 
\ee
Therefore
\be
  \widetilde S_h = S_\mathrm{DBI} + \int d^5x\, \Phi'  \rho \ .
\ee

In order to write down the final expression for the Legendre transformed action we need to eliminate $\Phi'$ from the Lagrangian density by inverting~\eqref{rhohomdef}. When doing this one should recall that we are working in an expansion of the original DBI action at small amplitudes of $F^{(L/R)}$, i.e., at small $\alpha'$. Therefore we will consistently ignore all such terms (typically higher powers of $h$) which correspond to higher order terms in the expansion of the DBI action. Doing this gives us the expression 
\be\label{Phipsol}
 \Phi' = -\frac{G \rho}{V_\rho w^2 e^{-4A}\sqrt{1+\rho^2\left(V_\rho w e^{-2A}\right)^{-2}}}\left[1-\frac{6\kappa\tau^2e^{-2A}h^2+6w^2e^{-4A}h^4}{1+\rho^2\left(V_\rho w e^{-2A}\right)^{-2}}+\frac{3}{2}\frac{w^2e^{-4A}f (h')^2}{G^2}\right] 
\ee
and allows us to cast the final action in a relatively simple form:
\begin{align} \label{LTactionfinal}
 \widetilde S_h &= -\int d^5x\,V_\rho G \sqrt{1+\frac{\rho^2}{\left(V_\rho w e^{-2A}\right)^2}}\left[1+\frac{6w^2e^{-4A}h^4+6\kappa\tau^2e^{-2A}h^2}{1+\rho^2\left(V_\rho w e^{-2A}\right)^{-2}}+\frac{3}{2}\frac{w^2e^{-4A}f (h')^2}{G^2}\right] \,, &
\end{align}
where
\be
 G = \sqrt{1+f\kappa e^{-2A}\left(\tau'\right)^2} \ .
\ee

In order to find the value of $r_c$, we need to minimize $\widetilde S_h$ at fixed $\varrho$. In the numerical analysis carried out below, we will simply do this by numerically minimizing of the action (which is finite for our ansatz so that no regularization is needed) rather than by using the equilibrium conditions explicitly. It is however instructive to compute the conditions.

There are two contributions to the equilibrium condition for $r_c$. The first one arises from the variation of $h$ which is necessary as the relation~\eqref{rho0def} needs to remain satisfied. The second one arises from the discontinuity of the Lagrangian density.\footnote{Recall that due to our prescription that delta functions at $r=r_c$ are neglected the two pieces of the Lagrangian for $r<r_c$ and for $r>r_c$ need to be treated independently.} The variation of the action due to changing $r_c$ at the discontinuity needs to vanish. The condition can be written as
\be \label{rccond}
 \mathrm{Disc}\, \frac{\delta h}{\delta r_c} \frac{\partial\widetilde  L_h}{\partial h'} = \mathrm{Disc}\,\widetilde  L_h 
\ee
evaluated on-shell at $r=r_c$, i.e., substituting the regular solution for $h$. Here $\widetilde L_h$ is the Lagrangian density of the Legendre transformed action $ \widetilde S_h$. This condition is the analogue of the stability condition~\eqref{baryoncond} in the thin layer approximation.

The remaining condition is the analogue of the condition~\eqref{Phirbval} and therefore should fix the normalization of the Abelian gauge field $\Phi$. Similarly as in the case of~\eqref{Phirbval}, there are two ways to derive it. First, we may require that there are only boundary contributions to the variation of the action. 
Second, we may replace the thermal gas geometry by a tiny black hole and require that the baryon charges behind the horizon and at $r=r_c$ are at equilibrium. 

We discuss first the variation of the action. We may keep $r_c$ fixed because assuming~\eqref{rccond} its variation does not contribute. After this, the solution only depends on the parameter $\varrho$. It appears explicitly in the solution~\eqref{rhosolhom} and affects indirectly the profile of $h$. The explicit dependence on $\varrho$ evaluates to an integral of $\Phi'$ after using the relation~\eqref{rhohomdef}. The variation due to the change in $h$ also gives a localized term. We obtain
\be
 \delta \widetilde S_h = - \mu\, \delta \varrho + \Phi(r_c)\,\delta \varrho +\mathrm{Disc}\, \delta h\, \frac{\partial\widetilde  L_h}{\partial h'} \ .
\ee
The extra terms in this differential vanish if
\be \label{rho0cond}
 \Phi(r_c) = -\, \mathrm{Disc}\, \frac{\delta h}{\delta \varrho} \frac{\partial L_h}{\partial h'} = 
 6 M^3 N_c V_{f0}(\l) w(\l)^2 e^{-\tau^2} f e^{A} G^{-1} \sqrt{1+\frac{\rho^2}{\left(V_\rho w e^{-2A}\right)^2}}\, \mathrm{Disc}\, \frac{\delta h}{\delta \varrho} h' \ .
\ee

We then consider the second condition. Taking the amount of charge behind the horizon to be $\rho_h$, the solution~\eqref{rhosolhom} is modified to read
\be \label{rhosolhommod}
\rho = \left\{\begin{array}{lr}
               \varrho  + \frac{2N_c}{\pi^2} e^{-b\,\tau^2}h^3(1-2b\,\tau^2) \, , \qquad&(r<r_c) \\
                \rho_h + \frac{2N_c}{\pi^2} e^{-b\,\tau^2}h^3(1-2b\,\tau^2) \, , \qquad&(r>r_c)
              \end{array}\right.
\ee
with
\be \label{rhohdef}
 \varrho -\rho_h = \frac{2N_c}{\pi^2} e^{-b\,\tau(r_c)^2}(1-2b\,\tau(r_c)^2)\ \mathrm{Disc}\, h^3(r_c) \ .
\ee
We then require the variation of the action with respect to $\rho_h$ near $\rho_h=0$ to vanish. The contribution due to the explicit dependence on $\rho_h$ again evaluates to an integral of $\Phi'$. We obtain
\be\label{rhohcond}
 \Phi(r_c) = \, \mathrm{Disc}\, \frac{\delta h}{\delta \rho_h} \frac{\partial\widetilde L_h}{\partial h'} = 
 -6 M^3 N_c V_{f0}(\l) w(\l)^2 e^{-\tau^2} f e^{A} G^{-1} \sqrt{1+\frac{\rho^2}{\left(V_\rho w e^{-2A}\right)^2}}\, \mathrm{Disc}\, \frac{\delta h}{\delta \rho_h} h' \ .
\ee
The condition is therefore very similar to~\eqref{rho0cond}, but the sign is opposite. We notice that it is the difference $\varrho-\rho_h$ which sources the discontinuity of $h$. In the limit of small density, $\rho$ can be neglected in the equation of motion of $h$ and then $h$ depends on the density only through the discontinuity. Therefore. in this limit, the variations of $\varrho$ and $\rho_h$ have exactly opposite effect on $h$. Consequently, the two conditions are equivalent. 

At finite density, it appears that the conditions~\eqref{rho0cond} and~\eqref{rhohcond} differ. In the current setup we simply choose to satisfy the first condition, as we are anyhow neglecting backreaction of the charge and higher order corrections in $h$ which are important at larger charge densities. 
Therefore, within the range of consistency of our approach, the conditions are equivalent, and we will simply ignore the disagreement at high density.
Alternatively, the system could be stabilized toward exchanging charge between the bulk and $r =\infty$ by adding some extra charge at $r=r_c$ (and naturally also by adding it at $r=\infty$).

\subsection{Asymptotic behavior}

We then discuss the asymptotics of the field $h$. It is straightforward to solve the asymptotics in the UV. After inserting the standard UV behavior in the action~\eqref{LTactionfinal} and solving for $h$, we find the asymptotics typical for gauge fields: 
$h \simeq C_1 +C_2 r^2$. We require that non-Abelian sources vanish, and therefore $C_1=0$, but $C_2$ remains as a free parameter. 

The IR behavior is more complicated. First, we notice that for $b>1$ in~\eqref{rhosolhom} and because the tachyon diverges faster than $r$ in the IR, the factor $\rho/(V_\rho w e^{-2A})$ tends to zero in the IR unless $h$ diverges exponentially fast. We will assume that this is the case and analyze the resulting EoM, which is obtained after neglecting the terms involving $\rho^2$ in the action:
\be \label{hIReq}
 \frac{1}{G f V_\rho w^2 e^{-4A}}\frac{d}{dr}\left(\frac{f V_\rho w^2 e^{-4A} h'}{G }\right) = \frac{4 h}{f w^2 e^{-2A}}\left(2 h^2 w^2 e^{-2A} + \kappa \tau^2 \right) \ .
\ee
First of all, $h=0$ is an exact solution to the full action. There is also a family of regular solutions
\be
 h \sim h_0 \exp\left[-C_\tau \tau(r)^2/r^2\right] \ .
\ee
which vanish in the IR. Here $C_\tau$ is a positive constant which can be expressed in terms of the IR expansion parameters of the potentials.
One may check numerically that there are no other asymptotic solutions (for which tachyon would diverge so fast that the condition of $\rho/(V_\rho w e^{-2A})$ vanishing in the IR would be violated).

The IR solutions in principle bring in one more parameter $h_0$ which also needs to be determined by minimizing the action. We have however checked that in the numerical analysis below the value of $h_0$ at the minimum is consistent with zero, i.e., in practice the solution in the IR region ($r>r_c$) simply vanishes.

\subsection{Phase diagram and the equation of state} \label{sec:homogeneousresults}

We work in the probe limit: we first construct the TG background solutions~\cite{Jarvinen:2011qe} in the absence of baryons, setting the quark mass to zero. The equations of motion and boundary conditions for the background fields $A$, $f$, $\lambda$ and $\tau$ are the same as those described in Sec.~\ref{sec:pointlikeeom}\@. We then insert the background in~\eqref{LTactionfinal} and solve the resulting equations of motion for the baryon field. We explicitly integrate \eqref{LTactionfinal} to obtain the free energy in the canonical ensemble, where we subtract $\tilde S_h$ evaluated for $h = 0$\@.\footnote{This subtraction makes sure that the grand potential is exactly equal to the Legendre transform of the free energy, not just up to an additive constant.}
We then need to minimize this free energy over the free parameter that we left as boundary condition, namely $C_2$  as defined in the previous subsection.
Subsequently, we do a Legendre transform to obtain the grand potential, which is used to compute the phase diagram, and which is equal to minus the pressure.
Note that in all the subsequent results, we set $N_c = 3$\@.

The precise choice of the various potential functions which we use in this section is given in Appendix~\ref{app:potentials}. Notice in particular that we choose the asymptotics of the $w$-function such that $w \sim \l^{-4/3}$ in the IR, with the explicit choice of the function given in~\eqref{eq:wpothomogeneous}. As we pointed out above in Sec.~\ref{sec:pointlikelocation} and proved in Appendix~\ref{app:wasympt}, we must choose the asymptotics $w \sim \l^{-w_p}$ with $w_p \ge 4/3$ in order to have a stable TG phase at small, nonzero $T$ and $\mu$, which is in turn necessary for the approach of this section to work. Moreover notice that in practice it is impossible to use the same sets of potentials in the two approximation schemes considered in this article: in the thin layer approximation of Sec.~\ref{sec:pointlike} we were forced to use $w \sim \l^{-w_p}$ with $w_p \le 2/3$ in order to stabilize the baryon phases.   

In the final results, we also compare the free energy of the TG and baryonic TG phases to the chirally symmetric, baryonless BH solutions at $x_f=1$, which are constructed as in~\cite{Alho:2013hsa}\@. The latter solutions therefore model the QGP, and the equation of state in this phase is that anchored to lattice data in~\cite{Jokela:2018ers} since the choice of potentials is the same. The comparison of the free energies between the baryonic and QGP phases however comes with uncertainties, which mostly arise from the normalization of the baryonic free energy. This is due to the approximations done in this section and the fact that the baryonic ansatz assumes $SU(2)$ flavor symmetry whereas the QGP equation of state was fitted to data with 2+1 dynamical quarks. In particular our result for the pressure in the baryonic phase may be somewhat low because of the simple approach.

\begin{figure}[t]
\centering
\includegraphics[width=0.9\textwidth]{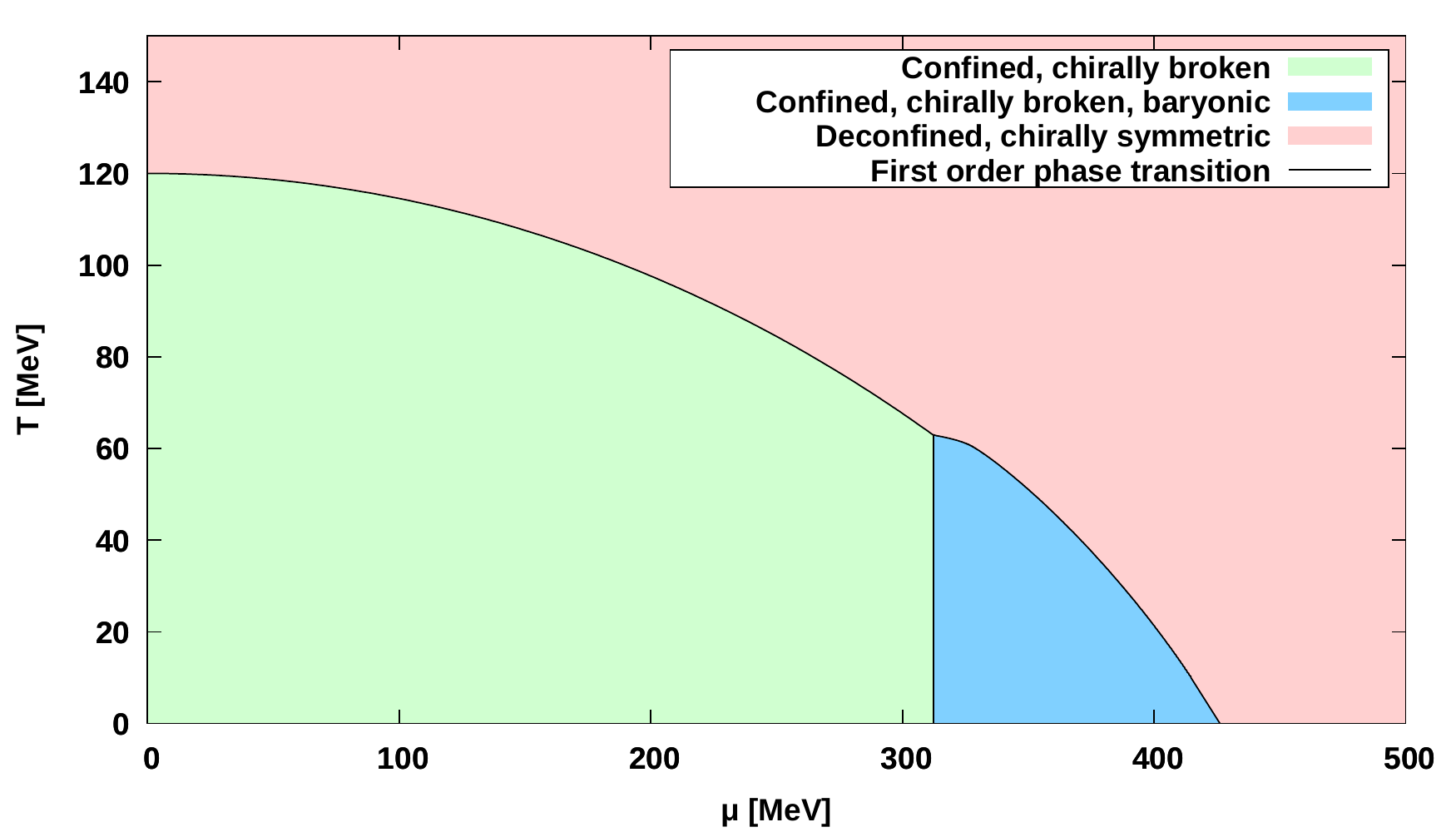}
\caption{\label{fig:smearedphases}Phase diagram on the $(\mu,T)$-plane for the homogeneous ansatz.}
\end{figure}

The phase diagram one obtains 
is shown in Fig.~\ref{fig:smearedphases}\@.
It features the following three phases, corresponding to the three kinds of numerical solutions we discussed above:
\begin{itemize}
\item A confining phase with broken chiral symmetry.
In contrast to the confining phase obtained in the thin layer approximation, this phase is a thermal gas phase, which implies that, for instance, glueballs are absolutely stable, not just long-lived like they were in the thin layer approximation.
\item A confining phase with broken chiral symmetry and condensed baryons, i.e., the new ingredient from the approach of this section.
\item A deconfined QGP phase with chiral symmetry.
\end{itemize}
The pressure in the first phase in our approach is constant~\cite{Alho:2013hsa}, and it is independent of the temperature also in the baryonic phase. Therefore the baryon-vacuum transition line is exactly vertical. Temperature dependence in these phases would arise from stringy loop effects~\cite{Alho:2015zua}. In the QGP phase, nontrivial temperature effects are included, and the result for the pressure can be seen as an extrapolation from the fit to lattice data around $\mu = 0$ as we have explained above~\cite{Alho:2013hsa,Jokela:2018ers}.

The location of this transition depends on the parameter $b$ described above.
In the results described in this section, $b = 10$ was chosen to have the transition at approximately $\mu = 313\,\text{MeV} \approx m_\text{proton} / 3$ (where we ignored the small binding energy of nuclear matter)\@.
Another thing to note is that, unlike within a similar approximation in the WSS model\footnote{A phase transition at a large value of the chemical potential is obtained in the WSS model in approaches using interacting solitons~\cite{Li:2015uea,BitaghsirFadafan:2018uzs}.}~\cite{Li:2015uea}, the baryonic phase does not survive to arbitrarily large values of the chemical potential.
Lastly, because in the thermal gas phase temperature dependence is suppressed, the baryonic phase transition does not end at a critical point, but instead continues until the QGP phase becomes dominant.

\begin{figure}[t]
\centering
\includegraphics[width=0.9\textwidth]{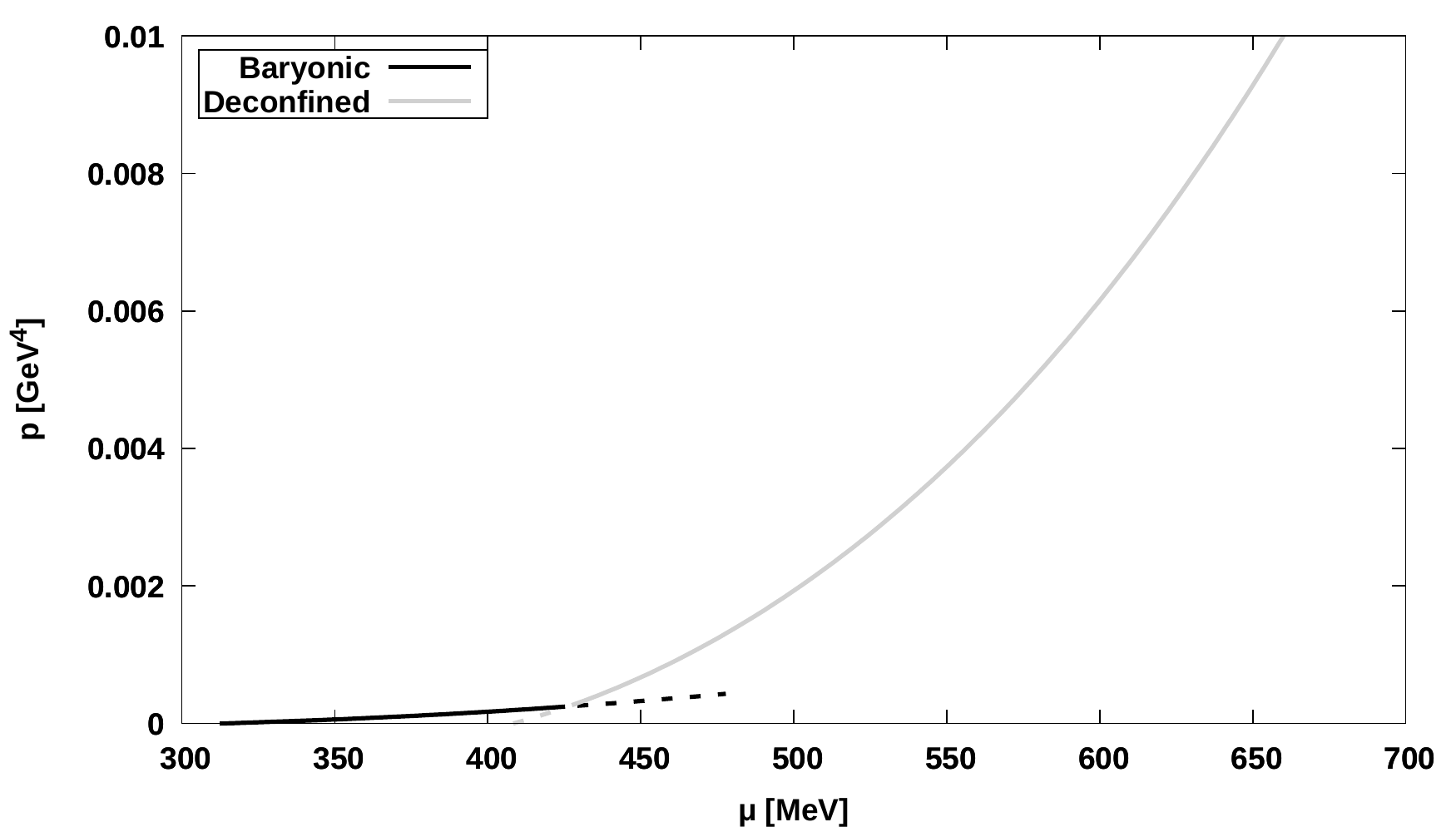}
\caption{\label{fig:pressure}Pressure at $T = 0$ using the homogeneous ansatz for the baryon phase. Metastable branches are denoted by dashed lines.}
\end{figure}

In Fig.~\ref{fig:pressure}, the pressure at $T = 0$ is plotted. 
The phase transitions are clearly visible. 
The latent heat at the vacuum to nuclear matter transition is $\Delta \epsilon \approx 51\ \mathrm{MeV}\, \mathrm{fm}^{-3}$, and at the nuclear matter to quark matter transition $\Delta \epsilon \approx 687\ \mathrm{MeV}\, \mathrm{fm}^{-3}$\@.
In principle, this pressure could be used as an equation of state to solve the Tolman-Oppenheimer-Volkov equations, thereby obtaining a mass-radius relation for neutron stars. (See, for instance, \cite{Kim:2011da,Ghoroku:2013gja} for the application in the WSS model.)
We leave this for future work.

\begin{figure}[t]
\centering
\includegraphics[width=0.9\textwidth]{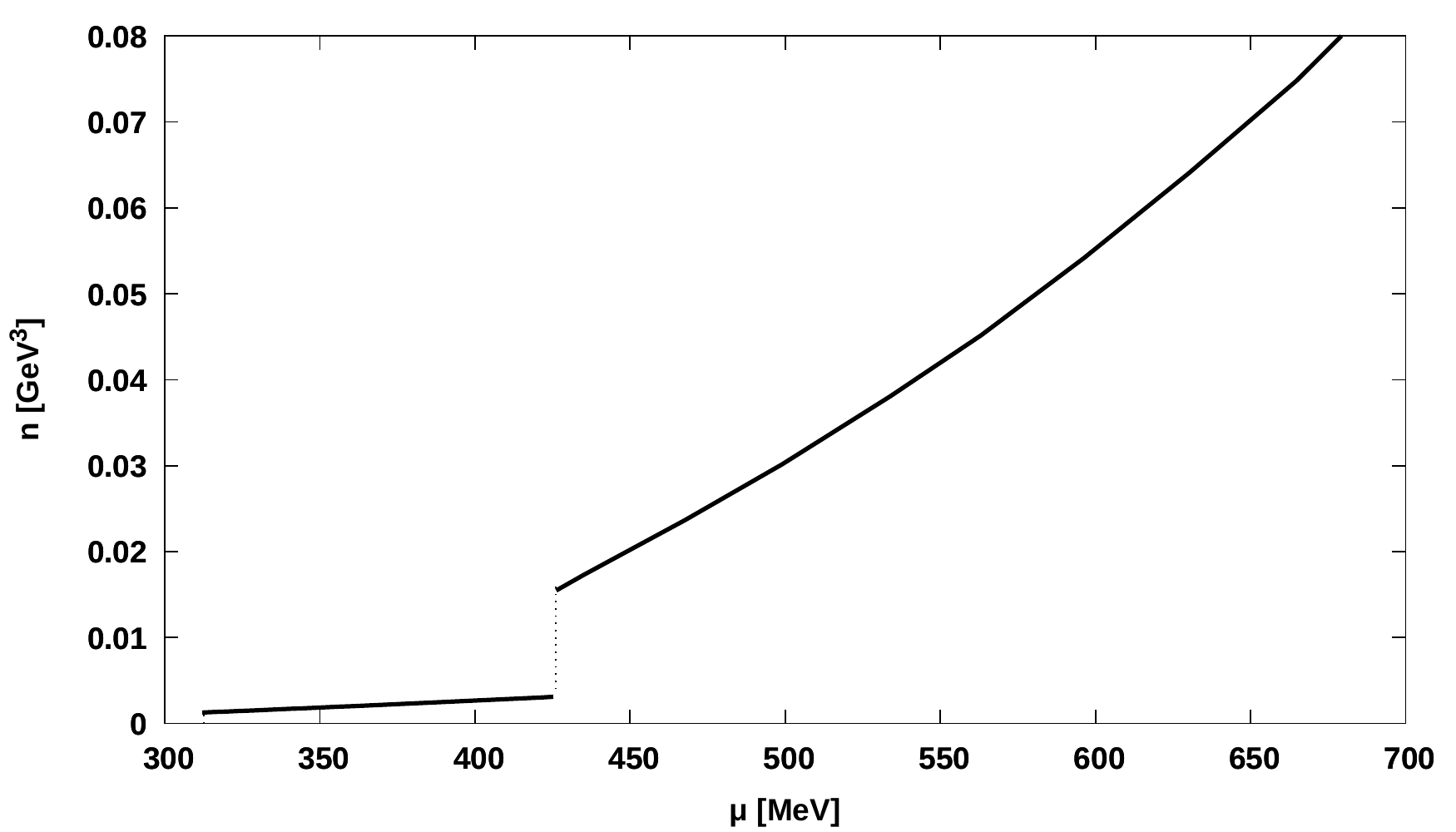}
\caption{\label{fig:numberdensity}Quark number density at $T = 0$  using the homogeneous ansatz for the baryon phase.}
\end{figure}

The quark number density is shown in Fig.~\ref{fig:numberdensity}\@.
From the fact that the number density jumps across the phase transitions, one can clearly see that the phase transitions are indeed first order. The baryon number density (which is obtained from the quark number density by dividing by $N_c=3$) at the end of the transition from the vacuum to nuclear matter is $n_b \approx 0.056\ \mathrm{fm}^{-3}$, i.e., $n_b \approx 0.35 n_s$ where $n_s \approx 0.16 \ \mathrm{fm}^{-3} $ is the nuclear saturation density.
This indicates that the nuclear matter number densities and pressures as a function of the chemical potential lie below the band of nuclear matter EoS (obtained by extrapolating the low density equations of state to higher densities~\cite{Hebeler:2013nza,Annala:2017llu}). 
This may be at least in part due to the uncertainties in the normalization of the baryon action~\eqref{LTactionfinal} and in the rough approximations done in this section. We comment more on this in Sec.~\ref{sec:conclusions}. We also note that the baryon to quark matter transition is strongly first order: the density jumps at the transition roughly by a factor $4.5$\@.

\begin{figure}[t]
\centering
\includegraphics[width=0.9\textwidth]{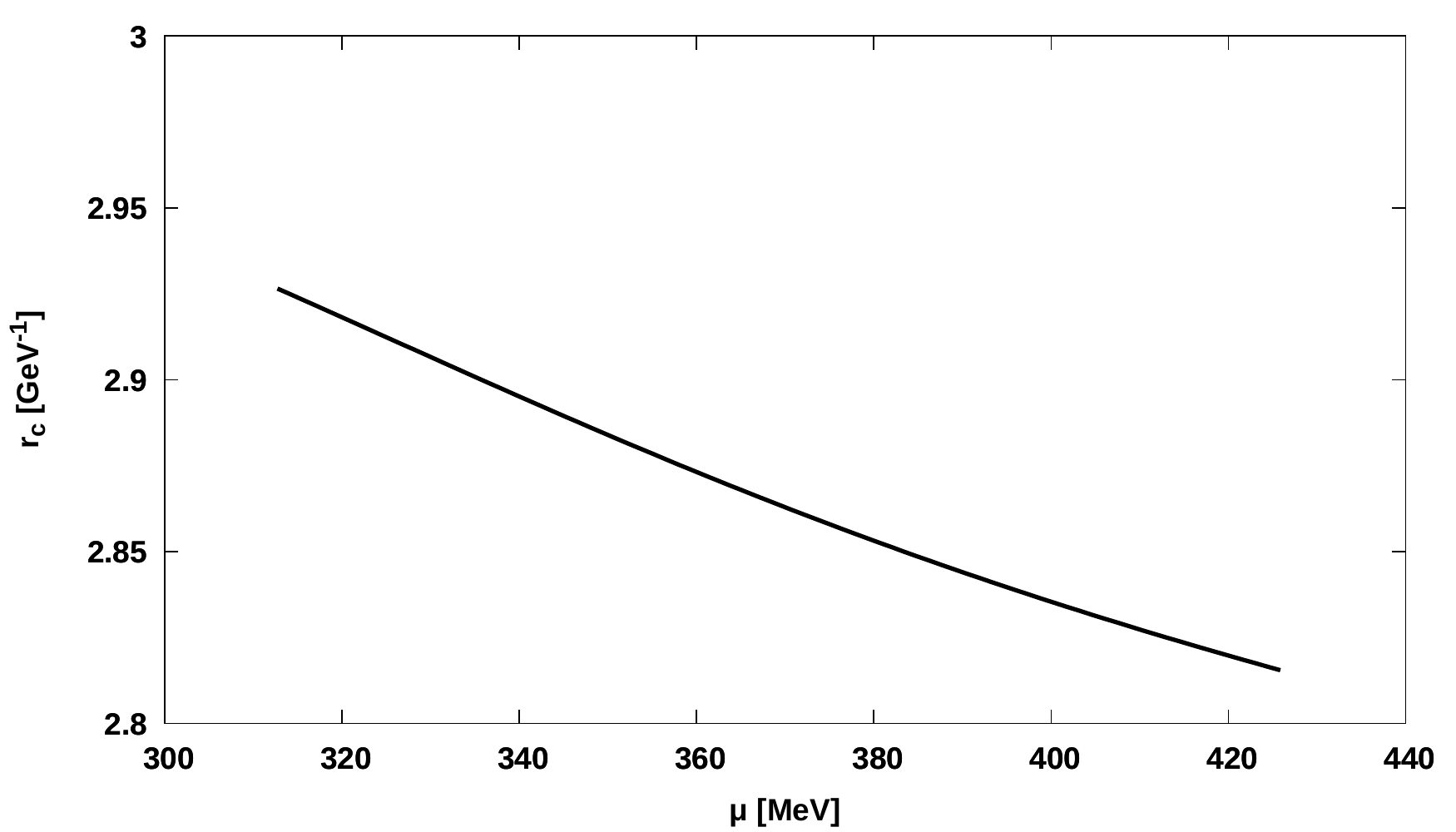}
\caption{\label{fig:smearedrc}Location of the soliton in the bulk at $T=0$ for the homogeneous ansatz.}
\end{figure}

Figure \ref{fig:smearedrc} shows the location of the discontinuity in the bulk.
The soliton stays roughly in the same place for different values of the chemical potential.
Also, in units of $\Lambda_\text{QCD}$, $r_c$ is roughly $\mathcal{O}(1)$, as one would expect.

\begin{figure}[t]
\centering
\includegraphics[width=0.9\textwidth]{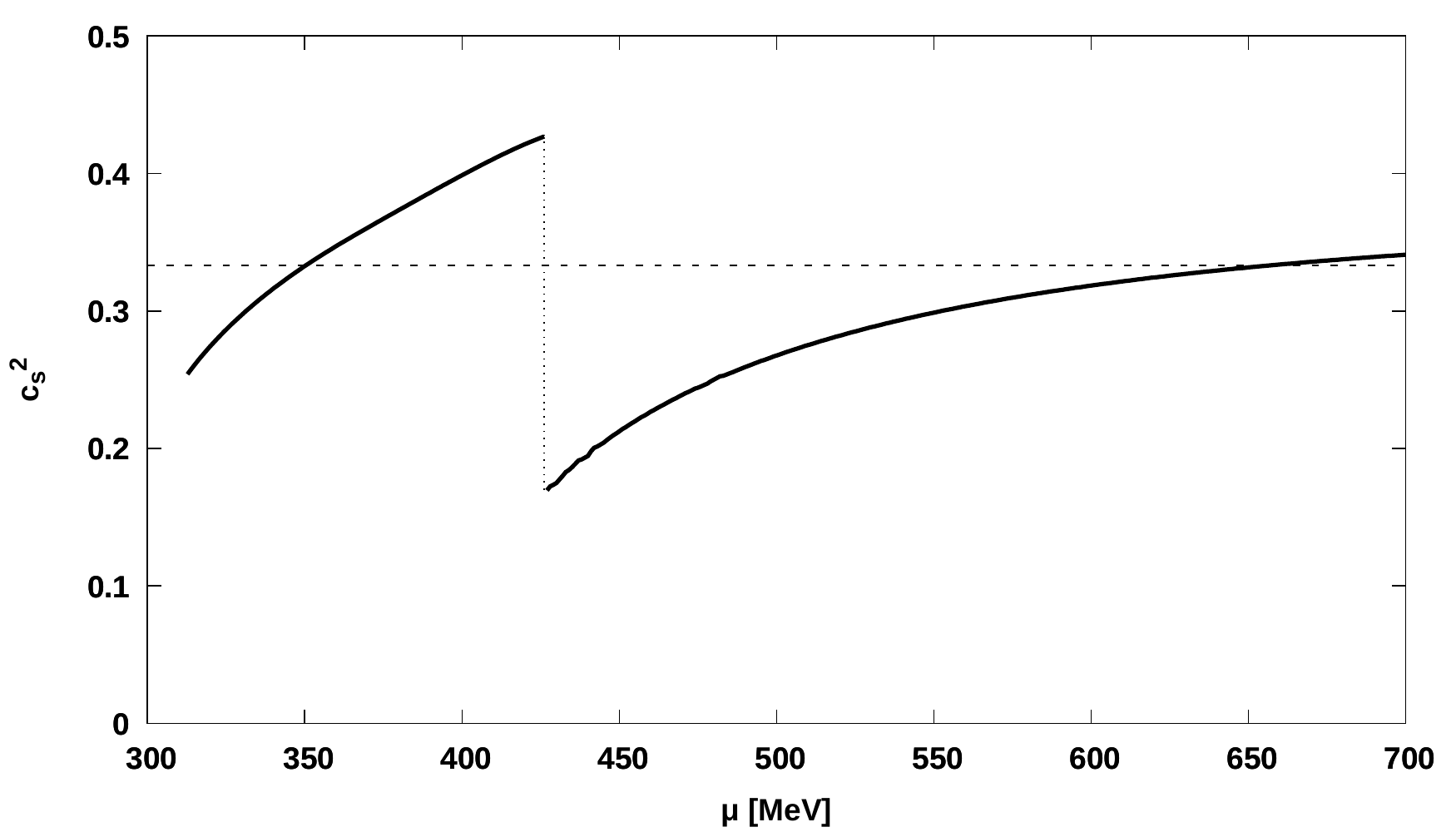}
\caption{\label{fig:smearedcssq}Speed of sound at $T = 0$. The dashed horizontal line denotes the conformal value.}
\end{figure}

The isothermal speed of sound is shown in Fig.~\ref{fig:smearedcssq}\@.
The vacuum (TG) phase  does not have a speed of sound, as it has no pressure and no energy density. At the vacuum to baryon matter transition the speed of sound immediately jumps to a value, which is larger than expected for nuclear matter in this regime. Such a deviation is not surprising because the homogeneous approximation used here is expected to be reliable only at higher densities.  The change of the speed of sound at the baryon to quark matter transition is relatively large, indicating a jump from a hard to soft equation of state at the strongly first order transition.
Noteworthy is that the speed of sound is above the conformal value $c_s^2 = 1/3$ in two intervals of the chemical potentials. 
For large values of the chemical potential, it approaches the conformal value from above. Notice that it is likely that speeds of sound above the conformal value are necessary in order to pass the astrophysical constraints from observations of neutron stars and their mergers (see, e.g.,~\cite{Bedaque:2014sqa,Tews:2018kmu}). Interestingly, the baryonic equation of state in the WSS model has been seen to also violate the conformal bound clearly~\cite{BitaghsirFadafan:2018uzs}. For other work towards realizing stiff phases in holographic models for dense QCD see~\cite{Hoyos:2016cob,Ecker:2017fyh}.

\begin{figure}[t]
\centering
\includegraphics[width=0.9\textwidth]{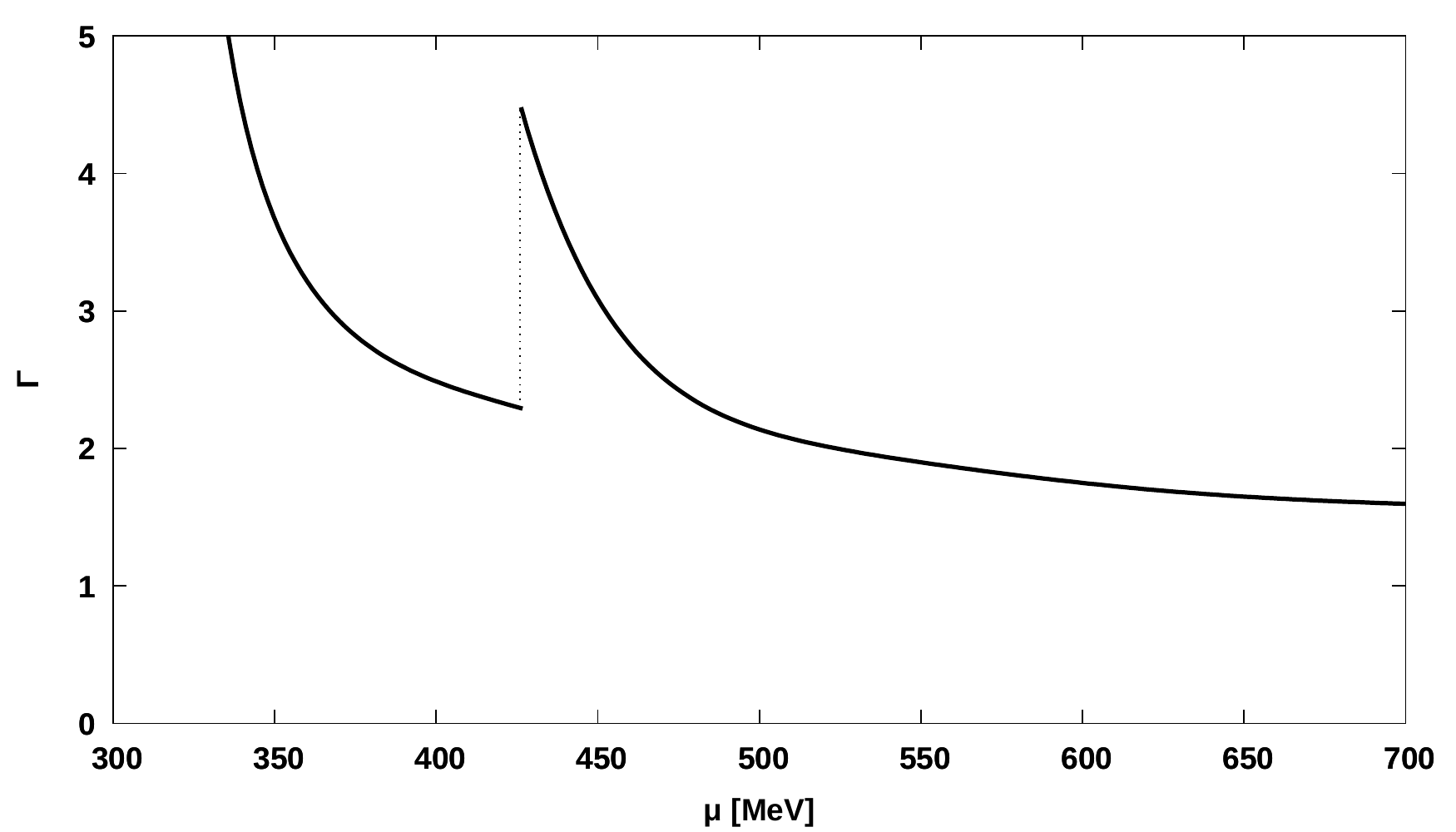}
\caption{\label{fig:smearedgamma}Adiabatic index at $T = 0$.}
\end{figure}

Lastly, the adiabatic index, which is defined as $\Gamma = \mathrm{d}\log p/\mathrm{d}\log n$, is plotted in Fig.~\ref{fig:smearedgamma}\@. Similarly as the speed of sound, this quantity measures the stiffness of the equation of state, and a piecewise constant $\Gamma$ is often used in polytropic realizations of the equation of state.
Its order of magnitude in the baryonic phase is comparable with what is expected from nuclear matter models (see, for example,~\cite{Douchin:2001sv,Hebeler:2013nza}).

\section{Discussion and Outlook}\label{sec:conclusions}

In this work, we explored how baryonic physics can be included in the V-QCD model.
Baryons play a large role in the cold dense matter region of the QCD phase diagram.
Neutron stars consist of matter located in this region of the phase diagram, which is one of the main motivations for this work.
In holography, baryons are dual to solitons sitting in the bulk.
Ideally, one would obtain these solitons explicitly.
This is unfortunately challenging especially in bottom-up models with potentials that are phenomenologically matched to QCD\@.
In this work, we tried to get around this problem by two different approximations: one in which the baryons are approximated as a thin layer of noninteracting solitons, and one in which they are described by a homogeneous non-Abelian gauge field configuration. These approximations are rough, and likely to miss some features of real baryon dynamics, but still we obtained several encouraging results.

Let us first discuss the thin layer approximation studied in Sec.~\ref{sec:pointlike}.
This approximation has baryonic matter appearing in roughly the expected region in the phase diagram in Fig.~\ref{fig:phases}.
Also, at large values of the chemical potential, the baryonic matter eventually makes way for a non-baryonic deconfined phase.
However, this approximation also has some unrealistic aspects, such as the appearance of a chirally symmetric baryonic phase.
The most serious of these issues is the stability of the location of the solitons.
In particular, the solitons fall into the deep IR unless in the asymptotic IR behavior of $w \sim \l^{-w_p}$, we choose $w_p \le 2/3$\@.
Such a power law has other consequences though, the most important of which is that at finite chemical potential the thermal gas is replaced by a small black hole phase which approaches a thermal gas solution as $\mu \rightarrow 0$\@.
This means that at finite chemical potential, the theory is only confining up to a certain distance scale, and that glueballs are no longer stable, but instead have a very long lifetime.

This issue is not present in the other approach we studied in Sec.~\ref{sec:homogeneous}, namely the homogeneous approximation.
In this approach, the solitons naturally stabilize at a coordinate distance from the boundary of order $\mathcal{O}(1)$\@.
In fact, as we discussed in Sec.~\ref{sec:backgroundthermo}, in order to satisfy phenomenological constraints the power law in the $w$ asymptotics has to be exactly equal to $-4/3$\@.

In the homogeneous approximation, the phase diagram (see Fig.~\ref{fig:smearedphases}) is qualitatively similar to what one would expect (within the limitations of the approach).
For small temperatures and small chemical potentials, there is a thermal gas phase, which displays confinement, a linear glueball spectrum and broken chiral symmetry.
At larger temperatures, there is a phase transition to a deconfined phase where chiral symmetry is restored.
If instead of increasing the temperature one increases chemical potential, a phase of non-zero baryon density appears, which is still confining and has broken chiral symmetry.
Increasing the chemical potential further, the baryonic phase gives way to a deconfined chirally symmetric plasma phase.

There is a hint that there is something going on in that region: at large chemical potential, as one takes $T \rightarrow 0$, the geometry becomes asymptotically $\text{AdS}_2$. This can be understood as a signal of a quantum critical regime~\cite{Alho:2013hsa}, which may be subject to instabilities. In this region the entropy is finite even at zero temperature. 
This could indicate that there is an operator missing in our bulk effective field theory, and the end point of the instability induced by this operator is another phase not included in the present analysis. It is tempting to interpret this as an instability towards an exotic, i.e., color superconducting phase in QCD. It would be interesting to understand if there are connections to the recent work on color superconductivity in other models~\cite{Faedo:2017aoe,BitaghsirFadafan:2018iqr,Faedo:2018fjw}.

In addition to the phase diagram, we computed several thermodynamical observables.
The most striking aspect of these is that the speed of sound clearly exceeds the value of conformal theories.
This result thus joins the list of examples where this happens~\cite{Hoyos:2016cob,Anabalon:2017eri,Ecker:2017fyh}.
From the point of view of neutron star physics, this result is not unexpected though, since experimental evidence seems to indicate that speeds of sound above the conformal value are necessary~\cite{Bedaque:2014sqa,Tews:2018kmu}.

For future work, it would be very interesting to apply our results to neutron star physics, following the ideas of~\cite{Hoyos:2016zke,Annala:2017tqz,Jokela:2018ers}. 
In particular, one could use the equation of state as input for the Tolman-Oppenheimer-Volkov equations to obtain a mass-radius relation for non-rotating neutron stars.
Also, one could use the equation of state to simulate neutron star mergers. Interestingly, our equation of state includes a first order transition between the baryon and quark matter phases at low densities which could be easily reachable in merger events. Actually, the densities which we obtained in this model are even too low to be consistent with extrapolations of equations of state from nuclear matter and the bound of the maximal neutron star mass from Shapiro delay measurements. This value is affected, among other things, by the normalization of the baryon action~\eqref{LTactionfinal}. This normalization contains uncertainties due to the inhomogeneities neglected in Sec.~\ref{sec:homogeneous}, and because we restricted to the ansatz with $SU(2)$ flavor symmetry which is also a rough model in the Veneziano limit. Improving on these issues may increase the pressure in the baryonic phase pushing the transition from baryons to quarks to higher chemical potentials. Note that such a change is also likely to make the excess over the conformal value for the speed of sound more drastic, see Fig.~\ref{fig:smearedcssq}. We could also simply treat the normalization of the action~\eqref{LTaction} as an additional fit parameter that could be determined e.g. by comparing to the nuclear saturation density as discussed in Sec.~\ref{sec:homogeneousresults}.

In light of neutron stars, it would also be interesting to expand the model to study the equation of state for neutron stars in more detail.
Currently, the V-QCD model that we are considering has a flavor sector consisting of $N_f$ identical massless quarks, all with charge $+1$\@.
In the interior of neutron stars, however, the number of neutrons is expected to be greater than the number of protons.
In principle, in the V-QCD model one could also differentiate between different quark species, giving each a different mass and couplings to external gauge fields.\footnote{Note that to incorporate the effect of multiple quark species onto the plasma phase, one would have to include the backreaction of each quark species. This would require that for each species $i$, there are $N_{f,i}$ identical copies.}
This would then allow to set the charges of the different species appropriately, as well as to introduce an isospin chemical potential, which could force an imbalance between the quark species to mimic that in a neutron star. 

One could also, since neutron stars are known to have strong magnetic fields, study the effect of a magnetic field on the baryonic matter.
Note that in \cite{Drwenski:2015sha,Gursoy:2016ofp,Gursoy:2017wzz}, the effects of a magnetic field on the plasma phase was already studied.
One could in principle then do a similar analysis as was done in this work to include the effect of the magnetic field on the baryonic matter too. This has been studied in the WSS model in~\cite{Preis:2011sp}. Another interesting extension would be to study the transport properties~\cite{Gursoy:2009kk,Iatrakis:2014txa} of dense nuclear matter. Dissipation plays a role even in isolated neutron stars, and (at least) bulk viscosity is expected to be relevant in neutron star mergers~\cite{Alford:2017rxf}.

Another thing that would be interesting is to explore different choices for the CP-odd potential.
In this work, we chose $V_a(\lambda, \tau) = \exp(-b\tau^2)$ with $b = 10$, but in principle one could even choose a different functional form entirely, possibly including dependence on $\l$. This will require establishing the CS terms of Sec.~\ref{sec:CS} for the more general choice of potential.
It would be interesting to study the effects of different choices on the baryon physics.
One could then try to match the potential to match nuclear matter models, or even the properties of neutron stars once more astronomical observations become available.

Another possible improvement to the approximations employed in this work would be to try to include the effect of backreaction of the homogeneous bulk gauge field $h$ onto the geometry.
This is in principle possible, but it complicates the numerical analysis, therefore we leave this for future work.
A similar possible improvement would be to include more terms in the DBI expansion using the results in the literature~\cite{Refolli:2001df,Koerber:2001uu,Grasso:2002wb,Keurentjes:2004tu}.
The full non-Abelian DBI action is not known, but one may use the action in~\eqref{generalact} with the standard  trace or by using the  symmetrized trace prescription. 
Moreover, one could check whether the related approximation scheme suggested in~\cite{Elliot-Ripley:2016uwb}, which is essentially based on applying the homogeneous assumption directly to the field strengths $F$ rather than the gauge fields, leads to any relevant changes in the results.

A final thing that would be interesting, if it can be done, is to attempt to obtain the soliton solutions explicitly.
This would 
certainly be worthwhile, for the following reason. 
As was mentioned in the beginning of Sec.~\ref{sec:homogeneous}, the homogeneous approximation implicitly assumes a high density of baryons.
This means that the approximation is not well suited to the low densities one expects near the vacuum to nuclear matter phase transition. In this regime description in terms of weakly interacting solitons could be more appropriate. If soliton interactions can be taken into account, they can then describe
both low densities and high densities equally well.

\section*{Acknowledgments}
We thank for discussions F.~Bigazzi, C.~Ecker, U.~G\"ursoy, C.~Hoyos, I.~Iatrakis, N.~Jokela, E.~Kiritsis, A.~Krikun, M.~Lippert, F.~Nitti, W.~van der Schee, A.~Schmitt, J.~Sonnenschein, and J.~Tarrio. We also thank C.~Ecker, N.~Jokela, E.~Kiritsis, W.~van der Schee, and A.~Schmitt for comments on the draft. This work is 
partially supported by the Netherlands Organisation for Scientific Research (NWO)
under the VIDI grant 680-47-518, and the Delta-Institute for Theoretical Physics ($\Delta$-ITP), both funded by the Dutch Ministry of Education, Culture and Science (OCW).
The work of T.~I.~is supported by JSPS KAKENHI Grant Number 18H01214.

\appendix

\section{Asymptotics of $w(\l)$ and the phase structure} \label{app:wasympt}

As it turns out the IR asymptotics of the coupling function $w(\l)$ of the gauge field may affect drastically the phase diagram of the model, which leads to a constraint for the IR behavior of this function which is completely independent of the presence of baryons. The solutions to analyze in order to see this are the small temperature near-extremal tachyonic black holes. In particular, as it turns out, it is important to compute what is the chemical potential in such backgrounds in the limit of small black holes. In this limit the chemical potential may either tend to zero or to infinity. The former option means that tachyonic black holes exist at arbitrary small chemical potentials and temperatures, and they will also dominate over the thermal gas solutions (which do not have a black hole horizon). Therefore the phase diagram has an undesired structure where the thermal gas phase (identified as the chirally broken confined phase of QCD in~\cite{Alho:2012mh,Alho:2013hsa}) is subdominant for all positive $\mu$. The dominant black holes are very small (as we shall see below) and the backgrounds are close to the thermal gas, so that the change in thermodynamics with respect to the thermal gas solution is tiny. However, the dominance of the black holes mean also that the Silver Blaze property of QCD (see~\cite{Armoni:2014wla}) is lost. As the result differs qualitatively from that obtained from QCD, the dominance of the black holes at small $T$ and $\mu$ is clearly disfavored.

We shall then analyze for which potentials this phenomenon occurs. Many of the details will be based on numerical analysis, and we will not present rigorous proofs. We assume that the IR behavior of the potentials is
\be
 V_g \sim \l^{4/3}\ , \qquad V_{f0} \sim \l^{v_p}\ , \qquad \kappa \sim \l^{-4/3}\ , \qquad w \sim \l^{-w_p}\ ,
\ee
up to logarithmic corrections in $\l$. The parameters must satisfy the following constraints: $4/3\le v_p\le 10/3$ and $w_p \le 4/3$~\cite{Jarvinen:2011qe,Arean:2013tja}. Further we will take the logarithmic corrections to $V_g$ and $\kappa$ to be those singled out in~\cite{Gursoy:2007cb,Gursoy:2007er,Jarvinen:2011qe,Arean:2013tja}. After this, our potentials satisfy all qualitative IR constraints established in earlier work. The potentials used in the numerical analysis of this article (given in Appendix~\ref{app:potentials}) belong to this class of potentials.

Let us first recall what is the ``standard'' behavior of the background in the IR in the absence of charge and black hole horizon, i.e., for the thermal gas solutions, for the specified potentials. The IR behavior in the gluon sector 
is given by~\cite{Gursoy:2007cb,Gursoy:2007er}
\be \label{Alaasympt}
 \log \lambda \sim \frac{3}{2}r^2 \ , \qquad A \sim -r^2\ , \qquad \left(\frac{r}{\Lambda} \gg 1\right) \ .
\ee
For the tachyon we have
\be \label{tauasympt}
 \tau \sim r^{C_\tau} 
\ee
where $C_\tau>1$ is a known constant. 

As we shall show, the structure of the relevant small BH backgrounds has the following scaling regimes: 
\begin{itemize}
 \item For $r \Lambda \ll 1$, the background follows the ``standard'' UV behavior established in~\cite{Gursoy:2007cb,Gursoy:2007er,Jarvinen:2011qe}. The UV regime will be irrelevant for the thermodynamics.
 \item For $r_* \Lambda \gg r \Lambda \gg 1$, where $r_*$ will be specified below, the background follows the ``standard'' IR behavior.
 \item For $r_h \gg r \gg r_*$, the metric will follow the standard IR behavior but the tachyon will be frozen. Here $r_h$ is the value of $r$ at the horizon.
\end{itemize}
The UV regime will be irrelevant for our analysis, so we restrict to the IR regime $r/\Lambda \gg 1$.  
The backreaction of the flavor to the glue is determined by the effective potential~\cite{Jarvinen:2011qe,Alho:2013hsa}:
\be
 V_\mathrm{eff}(\l,\tau) = V_g(\l) - x V_{f0}(\l)e^{-\tau^2} \sqrt{1+K} 
\ee
where
\be
 K = \left(\frac{\hat \rho}{e^{3A}V_{f0}(\l)e^{-\tau^2} w(\l)}\right)^2\ .
\ee
For the relevant solutions the charge $\hat \rho/\Lambda^3$ will be tiny so that for $r \sim 1/\Lambda$ the charge is decoupled, i.e., $K\ll 1$. Therefore we will be considering the limit $\hat \rho \to 0$. As $r$ grows, the solution behaves as the thermal gas and follows the asymptotics given in~\eqref{Alaasympt} and in~\eqref{tauasympt}. In particular the growth of the tachyon decouples the flavor from the glue. Because of the exponential dependence of $K$ on the tachyon, its growth will result in $K$ being of the order of one at some location, which we mark by $r_*$. Since the tachyon dependence dominates in $K$, we have roughly $\hat \rho \sim e^{-\tau(r_*)^2}$. Consequently,
\be \label{rstarIR}
 r_* \sim \left(-\log(\hat \rho)\right)^{1/(2C_\tau)} \ .
\ee
As $r$ grows further, $K$ will also grow. This is seen as follows. At large $K$ the tachyon EoM in~\eqref{taueq} becomes
\be
\frac{d}{dr}\left[ \frac{ f \kappa (\lambda)  \tau '}{w(\l)G}\right] \simeq 0 \ .
\ee
so that the term in the square brackets is a constant. We note that $f \kappa(\l)/w(\l)$ is suppressed in the IR: this is clear for $w_p<4/3$, and must also hold for $w_p=4/3$ for the spectrum to agree with QCD~\cite{Arean:2013tja}. But $\tau'/G$ is bounded  in no matter how fast the tachyon grows.
Therefore the constant must be zero, and the regular solution for the tachyon is simply the constant solution. Then using the asymptotics~\eqref{Alaasympt}, $K$ grows towards the IR if $w_p>v_p-2$. This equation is satisfied for the potentials we used in this article, and we will also assume this in the analysis below.\footnote{Our results will be valid also when this inequality is not satisfied, but the analysis will be more complicated.}

When $K \gg 1$, the effective potential reads
\be \label{VefflargeK}
 V_\mathrm{eff}(\l,\tau) \simeq  V_g(\l) -  \frac{x \hat \rho}{e^{3 A} w(\l)} \ .
\ee
Since $w_p>v_p-2>-2/3$, the second term will grow towards the IR and eventually become comparable to $V_g$. This marks the regime where the regular solution must develop a horizon: we may verify numerically that solutions with horizon at even higher values of $r$ do not exist. The solution in the vicinity of the horizon is not analytically tractable, and consequently we cannot estimate the temperature of the solutions analytically, but we may check numerically that as $r_h$ approaches its highest value $T/\Lambda$ tends to zero. From the requirement of the vanishing of~\eqref{VefflargeK} we obtain that
\be
  \frac{\hat \rho}{e^{3 A_h}} \sim \l_h^{4/3-w_p}\ , \qquad \hat \rho \sim \l_h^{-2/3-w_p} \ , \qquad r_h \sim \sqrt{\frac{-\log\hat\rho}{w_p+2/3}}\ .
\ee
This shows that we have the hierarchy assumed above, $r_h/ \Lambda \gg r_*/\Lambda \gg 1$, in the limit $\hat \rho \to 0$.

It remains to check the behavior of the chemical potential for these solutions. It is given by
\be \label{muint}
 \mu =-\int_0^{r_h} dr\, \Phi'(r)=\int_0^{r_h} dr\, \frac{\hat \rho}{V_{f0}(\l)e^{-\tau^2}w(\l)^2e^A}\frac{G}{\sqrt{1+K}} \ . 
\ee
The integrand $-\Phi'(r)$ increases fast with $r$ in the regime $r_* \Lambda \gg r \Lambda \gg 1$ due to the tachyon dependence. When $r_h \gg r \gg r_*$ we find that
\be
 -\Phi'(r) \sim \frac{Ge^{2A}}{w(\l)} \sim \l^{w_p-4/3} \ ,
\ee
where we used the fact that for the potentials and tachyon asymptotics specified above, $G$ has mild dependence on $r$ which we ignore here. For $w_p<4/3$ we find that $-\Phi'(r)$ decreases with $r$ in this region. Therefore the integral~\eqref{muint} is dominated at $r \sim r_*$, and we find
\be \label{muIR}
 \mu \sim  \l(r_*)^{w_p-4/3}
\ee
up to subdominant multiplicative corrections. For $w_p<4/3$ (but not for $w_p=4/3$) and given~\eqref{rstarIR}, we therefore find that $\mu\to 0$ as $\hat \rho \to 0$. Combining this with the numerical observation that these black hole solutions may also have an arbitrarily small temperature, we conclude that tiny tachyonic and charged black holes exist at arbitrarily small temperatures and chemical potentials.

Since $r_* \to \infty$ as $\hat \rho \to 0$, the background approaches  the thermal gas background pointwise in this limit. As the IR contributions to the grand potential are suppressed by the behavior of the metric and/or the tachyon, the value of grand potential for the black holes approaches that of the thermal gas. As $\hat\rho>0$ from the first law of thermodynamics it immediately follows that the black holes are dominant over the thermal gas solutions when $w_p<4/3$ (as was also seen numerically in Fig.~\ref{fig:phases}).

Therefore we conclude that a reasonable phase diagram can only be obtained for $w_p=4/3$: whether $\mu$ tends to zero or infinity in the zero charge limit depends on subleading corrections to~\eqref{muIR}. Numerically we have verified in this article and in~\cite{Alho:2013hsa} that for $w_p=4/3$ there are indeed choices of potentials which give the desired phase structure (i.e., no tachyonic black holes at small $T$ and $\mu$).

\section{Choice of potentials in the V-QCD action}\label{app:potentials}

We used two sets of potentials in this article, one set for the thin layer approximation in Sec.~\ref{sec:pointlike}, and another for the homogeneous baryons of Sec.~\ref{sec:homogeneous}. Notice that due to the constraints for the asymptotics of $w(\l)$ explained in Sec.~\ref{sec:pointlikelocation}, we cannot use the same set of potentials with both approaches. The sets of potentials were chosen to satisfy various asymptotic constraints (see, e.g.,~\cite{Arean:2013tja,Arean:2016hcs}) and fitted to lattice data for QCD thermodynamics as explained in~\cite{Jokela:2018ers}. The latter set equals the set 7a in~\cite{Jokela:2018ers} up to small changes in the potential $\kappa(\l)$ which leave the thermodynamics of the deconfined phase unaffected. For the former set, where we chose the function $w$ to have the asymptotics $w \sim\l^{-2/3}$ as we discussed in the main text, the fit is somewhat worse than for the latter set. That is, the lattice data favors the asymptotics $w\sim \l^{-4/3}$.

Notice that while the number of free parameters is high, the fit to lattice data is ``stiff'': after fixing the asymptotic behavior of the potentials, observables mostly depend very mildly on the details of the potentials in the regime $\l = \mathcal{O}(1)$. It is in this regime where most of the remaining freedom is, and it is controlled by the fit parameters discussed here.

For the potentials $V_g$, $V_{f0}$, and $\kappa$ we used the following ansatz in both cases:
\begin{align}
 V_g(\lambda)&=12\,\biggl[1+V_1 \l+{V_2\lambda^2
\over 1+\l/\l_0}+V_\mathrm{IR} e^{-\l_0/\l}(\l/\l_0)^{4/3}\sqrt{\log(1+\lambda/\l_0)}\biggr]\ ,& \\
 V_{f0}(\lambda) &= W_0 + W_1 \l +\frac{W_2 \l^2}{1+\l/\l_0} + W_\mathrm{IR} e^{-\l_0/\l}(\l/\l_0)^{2} \ ,& \\
\frac{1}{\kappa(\l)} &= \kappa_0 \left[1+ \kappa_1 \l + \bar \kappa_0 \left(1+\frac{\bar \kappa_1 \l_0}{\l} \right) e^{-\l_0/\l }\frac{(\l/\l_0)^{4/3}}{\sqrt{\log(1+\lambda/\l_0)}}\right] \ .&
\end{align}
Here the UV parameters were given by
\begin{align}
 V_1 &= \frac{11}{27\pi^2} \ , \qquad V_2 = \frac{4619}{46656 \pi ^4} \ , &\\
 \kappa_0 &= \frac{3}{2} - \frac{W_0}{8} \ , \qquad W_1 = \frac{8+3\, W_0}{9 \pi ^2} \ , \qquad W_2 = \frac{6488+999\, W_0}{15552 \pi ^4} \ ,&
\end{align}
where we set $x_f=N_f/N_c=1$, and chose $W_0=0$ for the first set (thin layer of noninteracting baryons) and $W_0=2.5$ for the second set (homogeneous field). The parameters for the glue potentials were chosen to be
\be
 \l_0 = 8\pi^2/3 \ , \qquad V_\mathrm{IR} = 2.05 
\ee
for both sets of potentials. They were obtained by fitting the lattice data~\cite{Panero:2009tv} for the thermodynamics of pure Yang-Mills theory~\cite{Alho:2015zua,Jokela:2018ers}.

The remaining parameters (mostly governing the potentials in the IR) and the function $w(\l)$ are different for the two potential sets.

The first set, used in Sec.~\ref{sec:pointlike}, has the following ansatz for $w(\l)$ 
\be
 \frac{1}{w(\l)} =  w_0\left[\sqrt{ \l/\l_0} + \frac{w_1 (\l/\l_0)^{3/2}}{1+\l/\l_0} + \bar w_0 (\l/\hat\l_0)^{2/3}\left(1+\frac{\bar w_1 \hat\l_0}{\l} \right) 
e^{-\hat \l_0/\l}\right]\ \label{eq:wpotpointlike},
\ee
where the UV behavior was chosen to produce qualitatively correct thermodynamics at large $T$ and small $\mu/T$~\cite{Jokela:2018ers}, and the IR asymptotics $w \sim \l^{-2/3}$ was chosen such that the baryon remains close to the UV boundary. The full set of UV and/or normalization parameters is
\be
 \kappa_1 = \frac{1}{3\pi^2}\ ,\qquad W_0=0 \ , \qquad w_0 = 0.93, \qquad w_1 = 0.75 \ , \qquad  M^3 = 1.12\ \frac{1+7/4}{45 \pi^2}\ .
\ee
Moreover, when $W_0=0$ the AdS radius $\ell =1$. The IR parameters are
\begin{align}
W_\mathrm{IR} &= 0.8 \ ,&  \bar\kappa_0 &= 1.5 \ ,&  \bar\kappa_1 &= -0.7\ ,& \\
\hat \l_0 &= 8 \pi^2 \ ,&  \bar w_0 &= 0.6 \ ,&  \bar w_1 &= 5.8 \ .&
\end{align}
These parameters were chosen such that the EoS at $\mu=0$ and its first nonzero cumulant compare relatively well with the lattice data. As we remarked above, due to the requirement $w \sim \l^{-2/3}$, the fit is a bit worse than for the potentials considered in~\cite{Jokela:2018ers} (and therefore also worse than for the other set given below).

The other set, used in Sec.~\ref{sec:homogeneous}, is the ``potentials 7a'' constructed in~\cite{Jokela:2018ers} up to small modifications in the choice of $\kappa(\l)$ which do not affect the thermodynamics of the deconfined quark matter phase. We choose 
\begin{align}
\frac{1}{w(\l)} &=  w_0\left[1 + 
\bar w_0 
e^{-\hat\l_0/\l}\frac{(\l/\hat\l_0)^{4/3}}{\log(1+\lambda/\hat\l_0)}\right]\ \label{eq:wpothomogeneous}. 
\end{align}
The UV parameters are given by
\be
 \kappa_1 = \frac{11}{24\pi^2}\ ,\qquad W_0=2.5 \ , \qquad w_0 = 1.28 \ , \qquad M^3 =  1.32\ \frac{1+7/4}{45 \pi^2\ell^3}
\ee
with the AdS radius $\ell = \left(1-2.5/12\right)^{-1/2}$.
The IR parameters are
\begin{align}
W_\mathrm{IR} &= 0.9 \ ,&  \bar\kappa_0 &= 1.8 \ ,&  \bar\kappa_1 = -0.23\ , \\
\hat \l_0 &= 8 \pi^2/1.18 \ ,&  \bar w_0 &= 18 \ . & &&
\end{align}

\section{Discontinuities and junction conditions for the thin layer approximation}\label{app:junctions}

In this Appendix we consider several technical details of the thin layer approximation of Sec.~\ref{sec:pointlike}. 

\subsection{Junction conditions}\label{app:jcs}

As we are backreacting the baryons to the metric, we 
need the Israel junction conditions for the bulk fields at the location of the baryon. We will consider here the general case where part of the charge sits behind the horizon and there is additional charge due to the baryon. That is, $\hat \rho= \hat \rho_h$ for $r>r_b$ and $\hat \rho = \hat \rho_h + \hat \rho_b$ for $0<r<r_b$. Since the source is independent of the tachyon, its junction condition is obtained by requiring the continuity of the term in the square brackets in~\eqref{taueq}. Denoting $f(r_b^\pm) = \lim_{r\to r_b\pm} f(r)$, we find
\be
 \tau'(r_b^-) =  \frac{\tau'(r_b^+)\sqrt{1+K(r_b^+)}}{\sqrt{1+K(r_b^+)+G(r_b^+)^2 (K(r_b^-)-K(r_b^+))}} \,,
\ee
where $K(r_b^-) = (\hat \rho_h + \hat \rho_b)^2 K(r_b^+)/\hat \rho_h^2$.
The source term however does depend on the metric and the dilaton, so its contribution to the Einstein equations needs to be considered. 
We notice that the dependence on the metric is through the factor $\sqrt{-g_{tt}}= \sqrt{f}e^A$. Consequently, the constraint equation~\eqref{ceq} is unchanged, but source terms are generated in the other equations:
\begin{align} \label{eeq1}
&\phantom{=}\ \ 6 f A''+6 f A'^2+3 f'A'+\frac{4 f\lambda '^2}{3 \lambda ^2}-e^{2 A}V(\lambda )+x e^{2 A}G \sqrt{1+K} V_f(\lambda ,\tau )&\nonumber\\
&= -(2\pi^2 M^3 \hat \rho_b )x V_{f0}(\l)w(\l)^2\sqrt{f}e^{-2A}\delta(r-r_b)\,, &\\
\label{eeq2}
&\phantom{=}\ \ f'' +3 f'A' - x e^{2 A}\frac{G K}{\sqrt{1+K}} V_f(\lambda ,\tau ) &\nonumber\\
&=(2\pi^2 M^3 \hat \rho_b )x V_{f0}(\l)w(\l)^2\sqrt{f}e^{-2A}\delta(r-r_b)\,.&
\end{align}
These equations imply the following discontinuities of the derivatives
\be
 A' = -\frac{1}{6 f} \mathcal{N}\theta(r-r_b)+\mathrm{continuous}\,,\qquad   f' =  \mathcal{N}\theta(r-r_b)+\mathrm{continuous} \,,
\ee
where
\be
 \mathcal{N} = (2\pi^2 M^3 \hat \rho_b )x V_{f0}(\l)w(\l)^2\sqrt{f}e^{-2A}\Big|_{r=r_b}\,.
\ee
The discontinuity of $\l$ can be read from its second order equation, which reads
\be
 \frac{f\l''}{\l^2} +\mathrm{regular} = \frac{3}{8} \mathcal{\widetilde N} \delta(r-r_b)\,,
\ee
where
\be
\mathcal{\widetilde N} = (2\pi^2 M^3 \hat \rho_b )x \sqrt{f}e^{-2A} \frac{d}{d\l}\left[V_{f0}(\l)w(\l)^2\right]_{r=r_b} \,.
\ee
This implies that
\be
 \l' = \frac{3\l^2}{8 f} \mathcal{\widetilde N} \theta(r-r_b)+\mathrm{continuous}\,.
\ee

\subsection{Baryon location} \label{app:baryonloc}

Having solved the discontinuities we then derive the equilibrium condition~\eqref{baryoncond}. In order to compute the variation of the first term in~\eqref{LTaction}, we first need to analyze the discontinuity of the tachyon more carefully. For clarity we replace the $\delta$ function by a continuous estimate $\delta_\e(r-r_b)$ and a continuous step function satisfying $\theta_\e'(r-r_b) = \delta_\e(r-r_b)$. Then we can write that
\be
 K = \frac{\left(\hat \rho_h+\hat \rho_b\, \theta_\e(r_b-r)\right)^2}{\left(e^{3A}V_f(\lambda ,\tau )w(\l)\right)^2} \ . 
\ee
However, the only property of $K$ that we will need below is the fact that near the discontinuity 
\be
 \frac{\partial K}{\partial r_b} \simeq -  \frac{\partial K}{\partial r}
\ee
since the terms where the derivative operates on the step function dominate.

The tachyon behavior near the discontinuity becomes
\begin{align} \label{taurbbeh}
 \tau'(r) &\simeq  \frac{\tau'(r_b^+)\sqrt{1+K(r_b^+)}}{\sqrt{1+K(r_b^+)+G(r_b^+)^2 (K(r)-K(r_b^+))}} \ ,& \\
  G(r) &\simeq G(r_b^+) \frac{\sqrt{1+K(r)}}{\sqrt{1+K(r_b^+)+G(r_b^+)^2 (K(r)-K(r_b^+))}} \ .&
\end{align}
The discontinuous behavior in the (derivative of the) first term of~\eqref{LTaction} is included in
\be
 G(r) \frac{d}{dr_b}\sqrt{1+K(r) } \simeq -\frac{1}{2}\frac{G(r)}{\sqrt{1+K(r) }}\frac{\partial K}{\partial r} \simeq -\frac{1}{2} \frac{G(r_b^+) }{\sqrt{1+K(r_b^+)+G(r_b^+)^2 (K(r)-K(r_b^+))}} \frac{\partial K}{\partial r}\ .
\ee
Notice that the discontinuity of the tachyon in $G(r)$ is a reaction to the explicit $r_b$ dependence of the charge density and the derivative should not act in this discontinuity. Integrating over $r$ yields
\begin{align}
&\phantom{=}\  \int dr\, G(r) \frac{d}{dr_b}\sqrt{1+K(r) }  &\nonumber\\
& \simeq   \frac{1}{G(r_b^+)} \left(\sqrt{1+K(r_b^+)}-\sqrt{1+K(r_b^+)+G(r_b^+)^2 (K(r_b^-)-K(r_b^+))}\right) \,. &
\end{align}

The second term in~\eqref{LTaction} can be treated in a more straightforward way, noticing first that the continuous estimates for the discontinuities of $\l'$, $A'$, and $f'$ are simply obtained by replacing $\delta \mapsto \delta_\e$ (which is the case because the coefficients of the $\delta$-functions in~\eqref{eeq1} and~\eqref{eeq2} are continuous). After partial integration, the contribution from the second term is
\begin{align} \label{baryonrhs}
 &\int dr\, \delta_\e(r-r_b) \frac{d}{dr}\left[V_{f0}(\l)w(\l)^2\sqrt{f}e^A\right] & \nonumber\\
= &\int dr\, \delta_\e(r-r_b)\sqrt{f}e^A \left[\l'\frac{d}{d\l}V_{f0}(\l)w(\l)^2 + V_{f0}(\l)w(\l)^2\left(A'+\frac{f'}{2f}\right) \right] \,. &
\end{align}
Taking into account the $\e$ regularization, one immediately obtains the expression in~\eqref{baryoncond} as the $r$-integral leads to the averages of the derivatives over the discontinuity.

As a consistency check, we show that the equilibrium condition follows from the consistency of the equations of motion. Namely, the constraint equation~\eqref{ceq} can be written as
\be \label{ceqreg}
 12 f A'^2 +3 f'A'-\frac{4  f \lambda '^2}{3 \lambda ^2}+ \frac{x e^{2 A}\sqrt{1+K} V_f(\lambda ,\tau )}{G} = \mathrm{continuous} \,,
\ee
where
\be
  \frac{\sqrt{1+K}}{G} \simeq \frac{\sqrt{1+K(r_b^+)+G(r_b^+)^2 (K(r)-K(r_b^+))}}{G(r_b^+)} \ .
\ee
Moreover, using
\be
 A'(r_b^\pm) = \langle A'\rangle \mp \frac{\mathcal{N}}{12 f}\,,\quad  f'(r_b^\pm) = \langle f'\rangle \pm \frac{\mathcal{N}}{2 } \,,\quad  \l'(r_b^\pm) = \langle \l'\rangle \pm \frac{3\l^2\mathcal{\widetilde N}}{16 f }
\ee
when evaluating the discontinuity of the left hand side in~\eqref{ceqreg}, we obtain
\begin{align}
 0 &= -\langle A'\rangle \mathcal{N} - \frac{1}{2f} \langle f'\rangle  \mathcal{N} - \langle \l'\rangle\mathcal{\widetilde N} & \nonumber\\
 &+x e^{2 A}V_f(\lambda ,\tau )\frac{1}{G(r_b^+)}\left[\sqrt{1+K(r_b^+)}-\sqrt{1+K(r_b^+)+G(r_b^+)^2 (K(r_b^-)-K(r_b^+))}\right] \,,&
\end{align}
which is the same condition as~\eqref{baryoncond}.

\subsection{Variation of the on-shell action}

Finally, we check explicitly that the contributions to the first law of thermodynamics from $r=r_b$ are absent. To this end we consider a generic variation of the on-shell action around a saddle-point configuration. As we already pointed out, the source term does not contribute to the variation thanks to the condition~\eqref{Phirbval}. As usual, the variation of the bulk term becomes a total derivative. In our case, the relevant variation terms read
\begin{align}
 \delta S_\mathrm{bulk} =\ &M^3 N_c^2 \int d^5 x \frac{d}{dr}\left[ \sqrt{-\mathrm{det} g} \left(g^{\m\n}\delta \Gamma^r_{\n\m} - g^{\m r}\delta \Gamma^\n_{\n\m} \right)\right] &\nonumber\\
 &+ \int d^5x\,\frac{d}{dr}\left[\frac{\pa\mathcal{L}_\mathrm{glue}}{\pa \l'} \delta \l\right]+ \int d^5x\,\frac{d}{dr}\left[\frac{\pa\mathcal{L}_\mathrm{DBI}^{(0)}}{\pa \Phi'} \delta \Phi\right]&\\
 =\ &- M^3 N_c^2 \int d^5 x\frac{d}{dr}\left[e^{3A}\left(5A'\delta f+8 f \delta A'+\delta f'+\frac{8f\l'\delta \l}{3\l^2}\right) + x \hat \rho \delta \Phi\right] \,,&
\end{align}
where we readily ignored the tachyon term which will not contribute at $r=r_b$. In order to make the variation of the gravitational action well-behaved we also need to add the Gibbons-Hawking term
\be
S_\mathrm{GH} = M^3 N_c^2 \int d^5 x\frac{d}{dr}\left[e^{3A}\left(8fA'+f'\right)\right] \ .
\ee
Adding the variation of this term, the result reads
\be \label{Sdisc}
 \delta S_\mathrm{disc} =  M^3 N_c^2 \int d^5 x\frac{d}{dr}\left[e^{3A}\left(24 f A'\delta A+3 f' \delta A+3 A'\delta f-\frac{8f\l'\delta \l}{3\l^2}\right) - x \hat \rho \delta \Phi\right] \,,
\ee
where the subscript ``disc'' refers to the fact that we are only keeping the terms which are potentially discontinuous at $r=r_b$. 

It suffices to show that the expression in the square brackets of~\eqref{Sdisc} is in fact continuous at $r=r_b$. To do this, we need to consider the variation of~\eqref{Phirbval}. This leads to two kind of contributions: one due to the variation of $r_b$ and the other due to the variation of the fields. The former was actually already computed above in Sec.~\ref{app:baryonloc} using the Legendre transformed action. In order to see this explicitly, we again consider a continuous estimate of the delta function, so that the variation of the left hand side of~\eqref{Phirbval} is interpreted as (omitting the trivial factor $\delta r_b$)
\be
 \int dr \delta_\e(r-r_b) \Phi'(r) \,.
\ee
Inserting here the solution of $\Phi'$ from~\eqref{rhodef} and $G(r)$ from~\eqref{taurbbeh} we find that
\begin{align}
  &\int dr \delta_\e(r-r_b) \Phi'(r)  \simeq \frac{1}{2 \hat\rho_b} \int dr \frac{e^{5A} V_f G(r_b^+)}{w(\l)\sqrt{1+K(r_b^+)+ G(r_b^+)^2(K(r)-K(r_b^+))}} &\\
  &\qquad = \frac{e^{5A}V_f(\l,\tau)}{\hat\rho_bG(r_b^+)}\left(\sqrt{1+K(r_b^+)}-\sqrt{1+K(r_b^+)+ G(r_b^+)^2(K(r_b^-)-K(r_b^+))}\right) \,. &
\end{align}
The variation of the right hand side of~\eqref{Phirbval} is computed as in~\eqref{baryonrhs} so the variations of $r_b$ cancel after imposing~\eqref{baryoncond}. Therefore we are left with the variations of the fields in~\eqref{Phirbval}, which lead to
\be
 e^{-3A} x \hat \rho_b \delta \Phi = \mathcal{N}\left(\delta A + \frac{\delta f}{2 f}\right)+\mathcal{\widetilde N}\delta \l \bigg|_{r=r_b} \ .
\ee
We then find for the discontinuity in~\eqref{Sdisc}
\begin{align}
 & \mathrm{Disc}\left[\left(24 f A'\delta A+3 f' \delta A+3 A'\delta f-\frac{8f\l'\delta \l}{3\l^2}\right) - e^{-3A} x \hat \rho \delta \Phi\right]_{r=r_b} &\nonumber\\
 =& - \mathcal{N} \delta A - \mathcal{N}\frac{\delta f}{2 f}-\mathcal{\widetilde N}\delta \l + e^{-3A} x \hat \rho_b \delta \Phi = 0 \,.
\end{align}
Collecting the results, we have shown that the variation of the on-shell action only receives contributions at the boundary.

As a final remark, we notice that the Gibbons-Hawking term 
\be
 S_\mathrm{GH} = M^3N_c^2\int d^4x\,\left[-L_\mathrm{GH}(r=0)+L_\mathrm{GH}(r_b^-)-L_\mathrm{GH}(r_b^+)\right] \ ,\qquad L_\mathrm{GH} = e^{3A}\left(8fA'+f'\right) \ ,
\ee
also contains a term localized at $r=r_b$ as $L_\mathrm{GH}$ is discontinuous at this point.

\bibliographystyle{JHEP}
\bibliography{refs}
\end{document}